\documentclass[titlepage]{amsart}
\usepackage{amsaddr}

\usepackage{productbox}
\usepackage{amsmath}
\usepackage{flexisym} 
\usepackage{breqn} 

\usepackage{amssymb}
\usepackage{amsthm}
\usepackage{amsfonts}
\usepackage{marginnote} 
\usepackage{mathrsfs} 
\usepackage[normalem]{ulem}
\usepackage[mathscr]{euscript} 
\usepackage{tcolorbox}
\usepackage{dcolumn}
\usepackage[T1]{fontenc}
\newcolumntype{2}{D{.}{}{2.0}}
\usepackage{scalerel}
\newcommand\scale[2]{\vstretch{#1}{\hstretch{#1}{#2}}}
\newcommand\ssp[2]{#1_{\scale{.8}{\scriptstyle #2}}}

\theoremstyle{definition} 
\newtheorem{theorem}{Theorem}
\newtheorem{lemma}[theorem]{Lemma}

\newtheorem{example}[theorem]{Example} 
\newtheorem*{remark}{Remark}
\newtheorem*{remarks}{Remarks}
\newtheorem{define}[theorem]{Definition}

\usepackage{empheq}

\usepackage{graphicx}
\usepackage{phaistos}
\usepackage{marvosym}
\usepackage{adforn}
\usepackage{tikzsymbols}
\usepackage{soul}
\usepackage{multicol}
\usepackage[]{cancel}
\usepackage{ifthen}
\usepackage[shortlabels]{enumitem}

\newcommand{\solution}{true} 

\ifthenelse{\equal{\solution}{true}}{\newcommand{\shs}[1]{{\noindent } #1}}
{}
\ifthenelse{\equal{\solution}{false}}{\newcommand{\shs}[1]{}
}{}

\theoremstyle{definition}

\newcommand{\df}[1]{{\bf#1}\index{#1}}

\makeindex 

\usepackage[pdfpagelabels]{hyperref}
\hypersetup{
    colorlinks,
    citecolor=blue,
    filecolor=blue,
    linkcolor=blue,
    urlcolor=blue
}

\usepackage{xcolor}     
\usepackage{lipsum}     

\usepackage{color,soul}
\usepackage{xcolor}
\usepackage{lipsum}

\definecolor{lightblue}{rgb}{.7,0.9,1}
\definecolor{forrestgreen}{rgb}{.1, .7, .1}
\usepackage{mdframed} 

\usepackage{mathtools} 

\newcommand{\curlnp}[1]{\text{ curl }#1}

\newcommand{\bz}{\mathbb{Z}}
\newcommand{\br}{\mathbb{R}}

\newcommand{\h}{h} 

\newcommand{\reflect}[1]{\textup{reflect}(#1)}
\newcommand{\In}[1]{\mathbf{In}(#1)}
\newcommand{\Out}[1]{\mathbf{Out}(#1)}
\newcommand{\hh}[1]{\mathbf{H}(#1)}
\newcommand{\om}[1]{\Omega(#1)}

\newcommand{\norm}[1]{\left\vert#1\right\vert}
\newcommand{\n}{\ensuremath{\mathbf{n}}\xspace}
\newcommand{\nn}{\ensuremath{\mathbf{N}}\xspace}

\newcommand{\p}{\mathbf{p}}
\newcommand{\QED}{\hfill {\Rectsteel}}
\newcommand{\q}{\mathbf{q}}
\newcommand{\xz}{\mathbf{x}}
\newcommand{\ry}{\mathbf{r}}

\newcommand{\ra}{\rightarrow}

\newcommand{\ray}[2]{[#1,#2]}

\newcommand{\set}[2]{\left\{#1 \mid #2 \right\}}

\newcommand{\sseq}{ \,\mathbb{\subseteq}\, }

\newcommand{\V}{\mathbf{V}}
\newcommand{\vv}{{\bf v}}

\newcommand{\x}{\mathbf{x}} 
\newcommand{\z}{\mathbf{0}}
\newcommand\scalemath[2]{\scalebox{#1}{\mbox{\ensuremath{\displaystyle #2}}}}

\title[First Integrals of Homogeneous Vector Fields and the Eigenmirror Problem]{First Integrals of Homogeneous Vector Fields and the Eigenmirror Problem of Geometric Optics}\thanks{This work is licensed under a Creative Commons Attribution-NonCommercial-NoDerivatives 4.0 International License (CC BY-NC-ND 4.0).}

\author[R. A. Hicks]{R. Andrew Hicks}
\address{Department of Mathematics, Drexel University, Philadelphia PA 19104} 
\email{ahicks@drexel.edu}
\subjclass[2020]{Primary 78A05; Secondary 34A26, 34C45}

\date{\today}

\begin{document}

\begin{abstract}
The eigenmirror problem asks: ``When does the reflection of a surface in a curved mirror appear undistorted to an observer?'' We call such a surface an {\bf  eigensurface} and the corresponding mirror an {\bf  eigenmirror}. The data for an eigenmirror problem consists of a homogeneous transformation ${\bf H}:\mathbb{R}^3 \to \mathbb{R}^3$ that encodes what it means for two observers to see a surface in the ``same way.'' 

A solution to this problem is a differentiable 2-manifold that (1) satisfies a first-order partial differential equation called the {\bf anti-eikonal equation}, and (2) satisfies certain side inequalities that ensure that a ray reflecting off the mirror behaves in a physically meaningful way. Although these side inequalities initially seem like an ad hoc global restriction, we show that under reasonable conditions, an integral curve of the characteristic flow of the anti-eikonal equation may not intersect the boundary of an eigenmirror. Thus, in those cases, the eigenmirror is invariant under the characteristic flow. We give several examples exhibiting our results.
\end{abstract}

\maketitle
\tableofcontents


\section{Introduction: The Eigenmirror Problem}

\begin{figure}[htbp]
\centering
\includegraphics[height=2in]{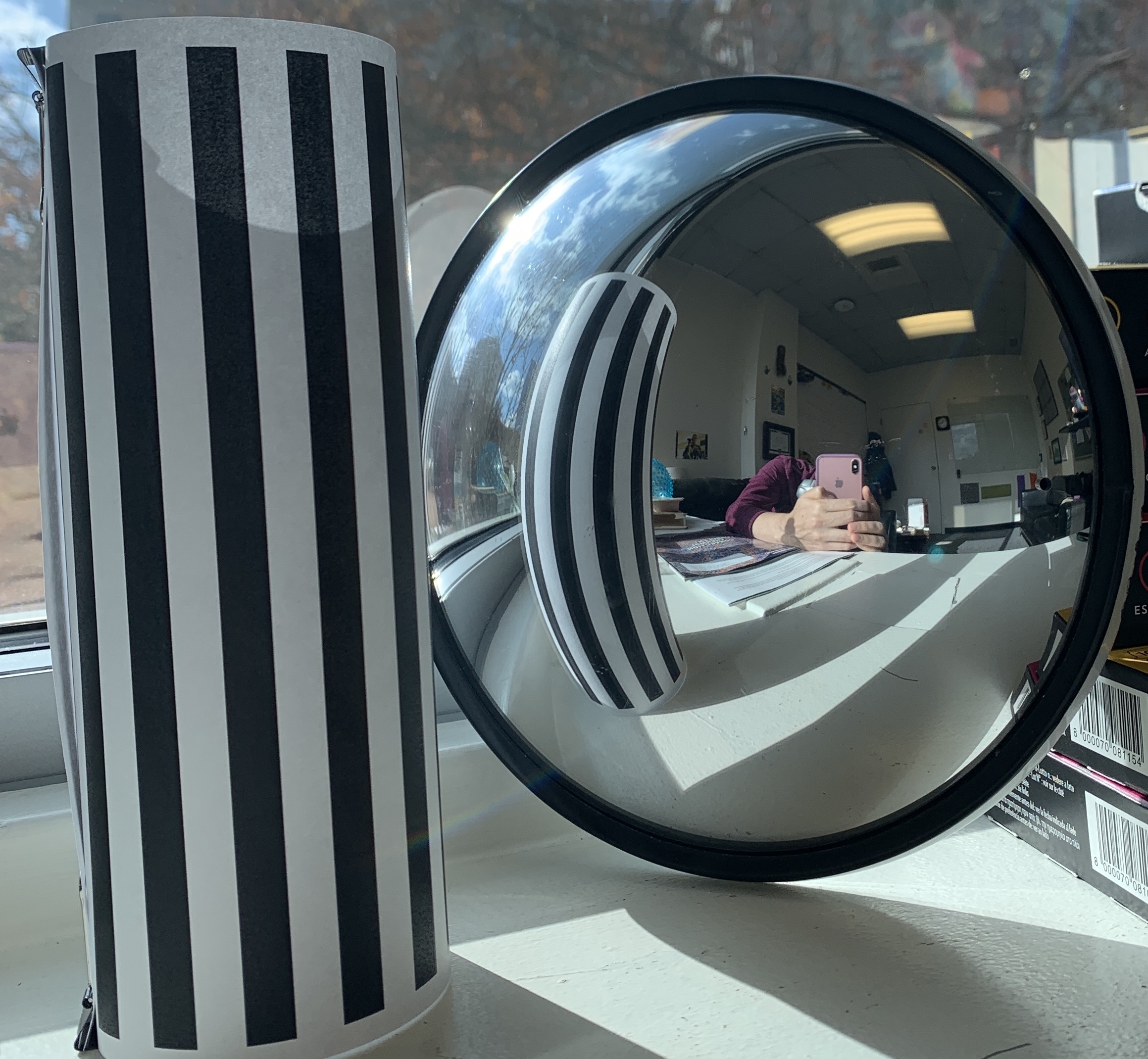}\qquad 
\includegraphics[height=2in]{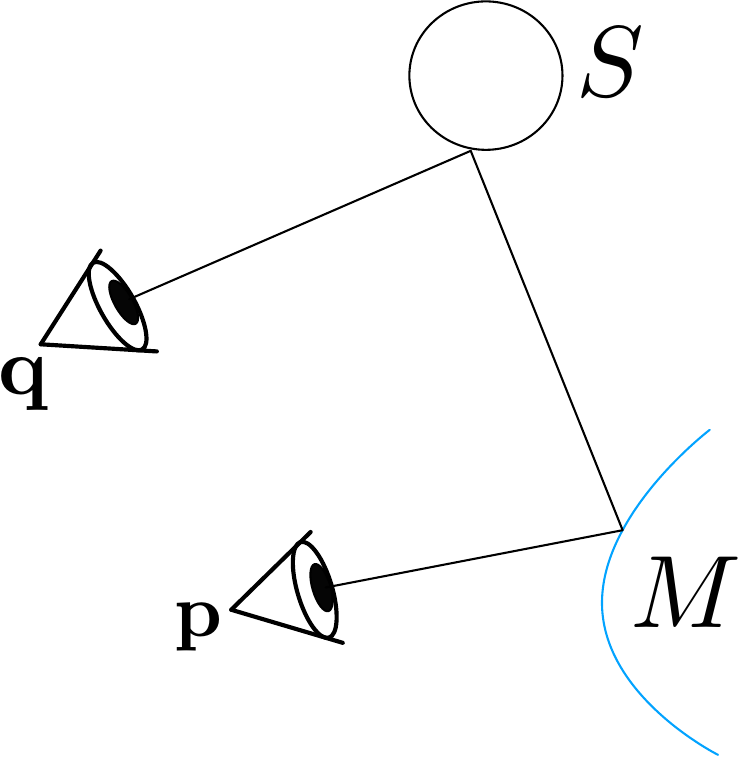}
\caption{On the left,  the reflection of a cylinder  in a spherical mirror. On the right, we see a  rough schematic of the general eigenmirror problem. Our observers have monocular vision, which functions as a pinhole camera.}
\label{fig:cylinder}
\end{figure}

When gazed at, most curved mirrors create distorted reflections. For example, consider the reflection of the striped cylinder in the spherical mirror depicted on the left of Fig. \ref{fig:cylinder}. 
The eigenmirror problem asks, roughly: If one views a surface $S$ in a reflector $M$, when does it appear undistorted?''  When $M$ and $S$ have this property, we call $S$ an \df{eigensurface}, and the mirror, $M$, the \df{eigenmirror}. We require that $M$ be a differentiable 2-manifold, and that the observer's and mirror's locations be fixed. 

The use of the term ``undistorted'' is what causes some difficulty in stating our problem. In what follows, let's identify observers with their locations. 

Initially, we want to state that, as shown on the right in Fig. \ref{fig:cylinder}, an observer $\p$ looking into $M$ sees a reflection of the surface $S$.
And that $S$ appears to $\p$ in this reflection just as $S$ appears to an observer $\q$ who looks directly at $S$.

$\p$ may or may not equal $\q$, but here we will mostly focus on the case of $\p=\q$.
How do we formalize this?

\begin{figure}[htbp]
\centering
\includegraphics[height=2in]{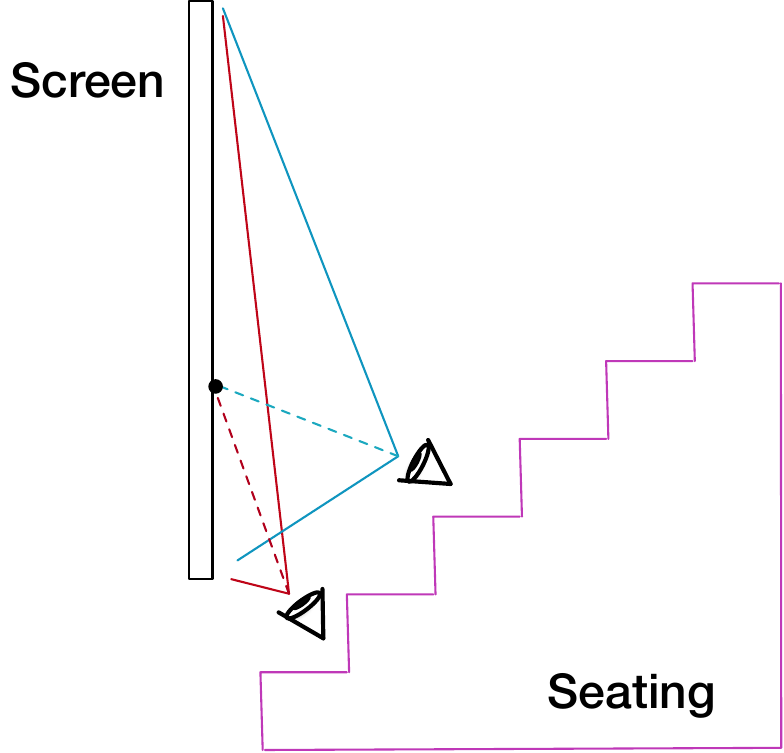}
\caption{Two friends go to a  movie, but one must sit close to the screen. We may consider the rays entering the eyes of the friends to be in 1-to-1 correspondence. For example, the two dotted rays are associated with each other.}
\label{fig:movie}
\end{figure}
Suppose two  mathematicians go to see a movie in a crowded theater. Only two seats are left: one in the front row, close to the screen, and another, much better seat, in a middle row.

The first mathematician, grumbling, takes the front row seat, while the second argues, ``It will be fine - you'll see the entire screen, just as I will, since there is a nice 1-to-1 correspondence, ${\bf H}$, between the sets of rays that will enter our eyes.''
 We will refer to sets of rays as \df{bundles of rays}, and for our movie watchers we mean that all the rays entering the eye of an observer go through a single fixed point.

The details don't matter, but the point is that the transformation between two ray bundles is determined in this situation. 

As in the movie theater example, the transformation is determined for the problem with a single mirror, as shown on the right of  Fig. \ref{fig:cylinder}. If that transformation is equal to a prescribed ${\bf H}$, then $M$ is an eigenmirror and $S$ an eigensurface, with respect to ${\bf H}$. 

\begin{figure}[t]
\begin{center}
\includegraphics[width=50mm]{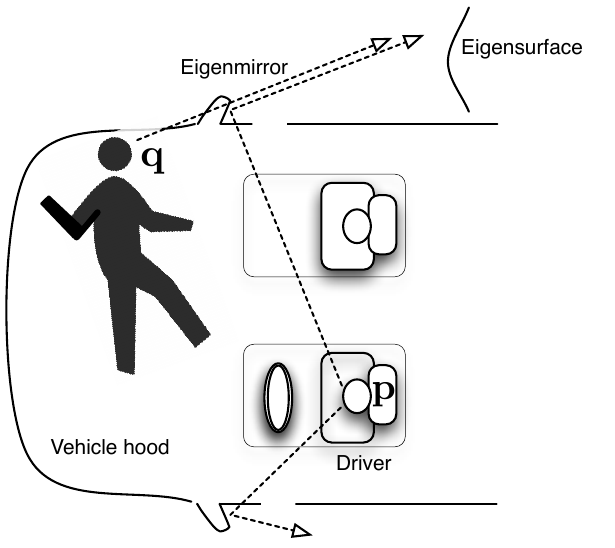} 
\caption{A passenger side mirror (US) may have a blindspot. Treating the surface of an oncoming car as $S$, the driver desires a sideview mirror $M$, so that their view of $S$ in the mirror is the same view that the person laying on the hood has.}
\label{fig:on-the-hood}
\end{center}
\end{figure}

{\bf History and Motivation}
This work arose from the related problems of panoramic vision for robot soccer \cite{hicks99cvpr} and the problem of blindspots in automotive mirrors \cite{hicks05ao-blind}. A driver with blind spots in their side-view mirrors seeks a wider, undistorted view. One way to model this is by saying that the driver, $\p$, desires to see the surface of an approaching car, $S$, in their mirror $M$, in the same way a person, $\q$, lying on the hood of their car would see $S$ - see Fig. \ref{fig:on-the-hood}.

A second source of motivation is the topic of \df{optical fabrication}, which largely refers to the creation of mirrors and lenses. The challenges are numerous, but one must get the geometry right and also polish these components to a degree of smoothness dictated by the application of interest. This history is long and complex, but traditionally, optical designers have been restricted to using mostly spherical or flat components. Of course the Newtonian reflector is a parabola but is well-approximated by a sphere. As a teenager taking a telescope making class, the author was told by his teacher that a parabola was a ``deepened'' sphere, and the class was instructed to grind the glass blanks a bit more in the center. 

Another historical example was the Schmidt telescope reflector plate, invented in 1930, was rotationally symmetric but aspherical \cite{kingslake-history}. 
Remarkably, the first successful product employing fully asymmetric surfaces was the Polaroid SX-70 Land camera, which debuted in 1972 and  took instant photos \cite{plummer82ao}. Elaborate methods needed to be invented to produce it, and the cost was \$100 million dollars at the time \cite{bonanos2012instant}. The current commercial landscape of asymmetric surfaces is vast - see \cite{rolland22survey}.\footnote{The optical community refers to asymmetric surfaces as \df{freeform surfaces}.}

Rotationally symmetric components can be made on a computer controlled lathe, but it wasn't until the 1980s that aspherical rotationally symmetric surfaces began to become economically feasible. (At the time of writing this paper aspheres are still more expensive.) However, surfaces that were not rotationally symmetric were practically unheard of since they could not be fabricated. As a result, no one was going to start looking at problems modeled with partial differential equations (PDEs) if the solutions were asymmetric.

However, in the late 1990s, technology became available that allowed for the optical fabrication of essentially any surface, polished to the smoothness of the best telescope mirrors. A lovely asymmetric mirror problem  that was solved around this time is the single-source illumination problem, which employs the Monge-Amp\'ere equation - see \cite{oliker97ip, oliker98num}. 
This was the beginning of the development of deep connections between optimal transport and geometric optics - see \cite{rubinstein17josaa, gutierrez2017na, oliker17opex}. 

Conic sections answer many optical questions, but many fundamental questions with polynomial answers have likely been left unexamined due to fabrication restrictions. As a result, there is little theory that addresses asymmetric surfaces. 

The idea of an eigenmirror was pointed out to the author by his colleague R. Perline over 20 years ago, but then put aside since the author found the problem difficult to formulate. Eventually, it was revisited in  \cite{esofem19josaa, hicks20josaa}. In some sense, the eigenmirror problem is like the rigid body problem of mechanics, where first integrals play an important role. For this reason, the author's thinking has been influenced by \cite{grammaticos90,llibre06,goriely2001}.

\section{Two Elementary Examples}

Throughout this paper, we will always treat rays as emanating {\em out} of the eye of observer. This is fine, because geometric optics is reversible.

\begin{example}

Let's consider the case of a flat mirror in the plane, as in Fig.~\ref{fig:virtual}A. An observer at $\p$ views a mirror $M$ and sees the reflection of $S$, while an observer at $\q$ views $S$ directly, ignoring any possible obstruction by $M$. We will refer to the observers as $\p$ and $\q$. Here, we will see that there is a flip between two ray bundles. 

In \ref{fig:virtual}B, we see a ray $\ry$ exiting $\p$. We denote  the corresponding ray from $\q$ by $\hh{\ry}$. In \ref{fig:virtual}C we have a different setting, where $\p$ and 
$\q$ have different fields of view. Thus ${\bf H}$ expands the bundle leaving $\p$.

Thus we see from these diagrams that there is a transformation ${\bf H}$ at work here. For both of these configurations $S$ is an eigensurface of $M$, given the appropriate ${\bf H}$.

So the notion of $\p$ and  $\q$ seeing the same thing is a bit subtle. Probably for humans taking ${\bf H}$ to be linear is reasonable (think of the movie theater), but the author is reluctant to commit. Instead of debating what the correct definition is of ``the same'', one can simply work under the assumption that  ${\bf H}$ is given.

\begin{figure}[htbp]
\centering
{\scriptsize A.}\includegraphics[height=1.2in]{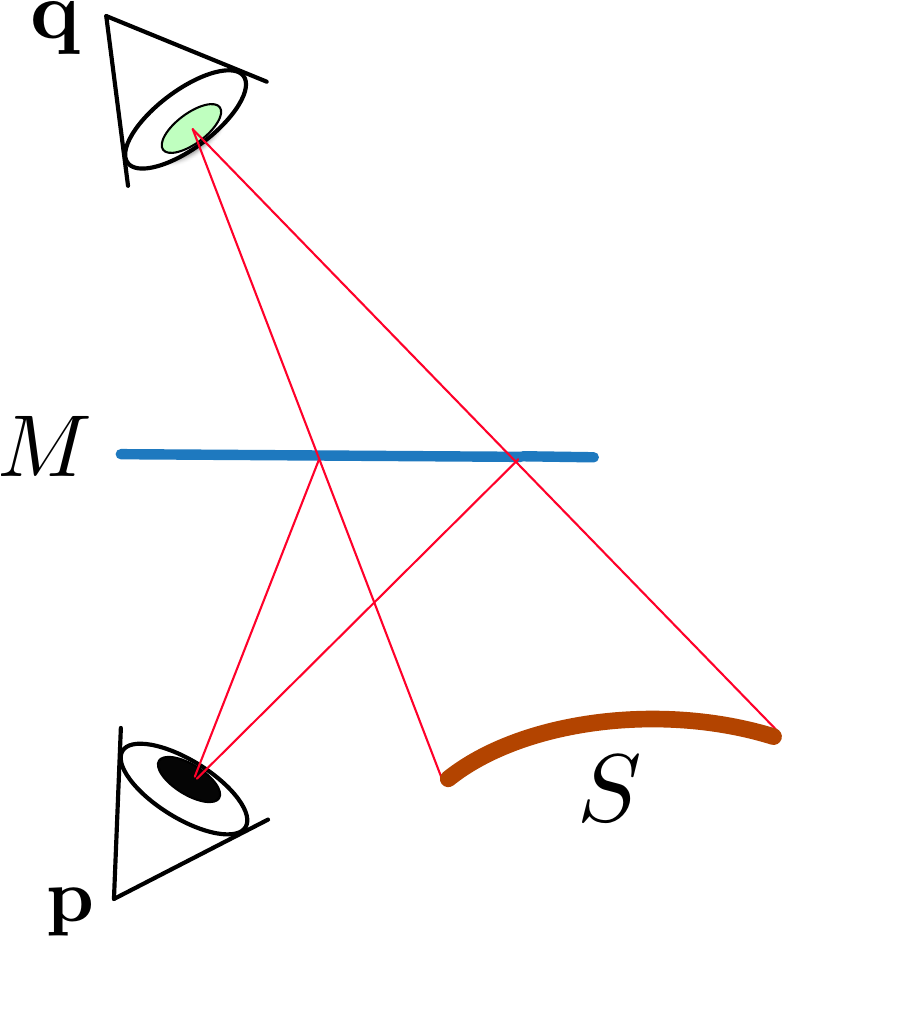}
{\scriptsize B.}\includegraphics[height=1.2in]{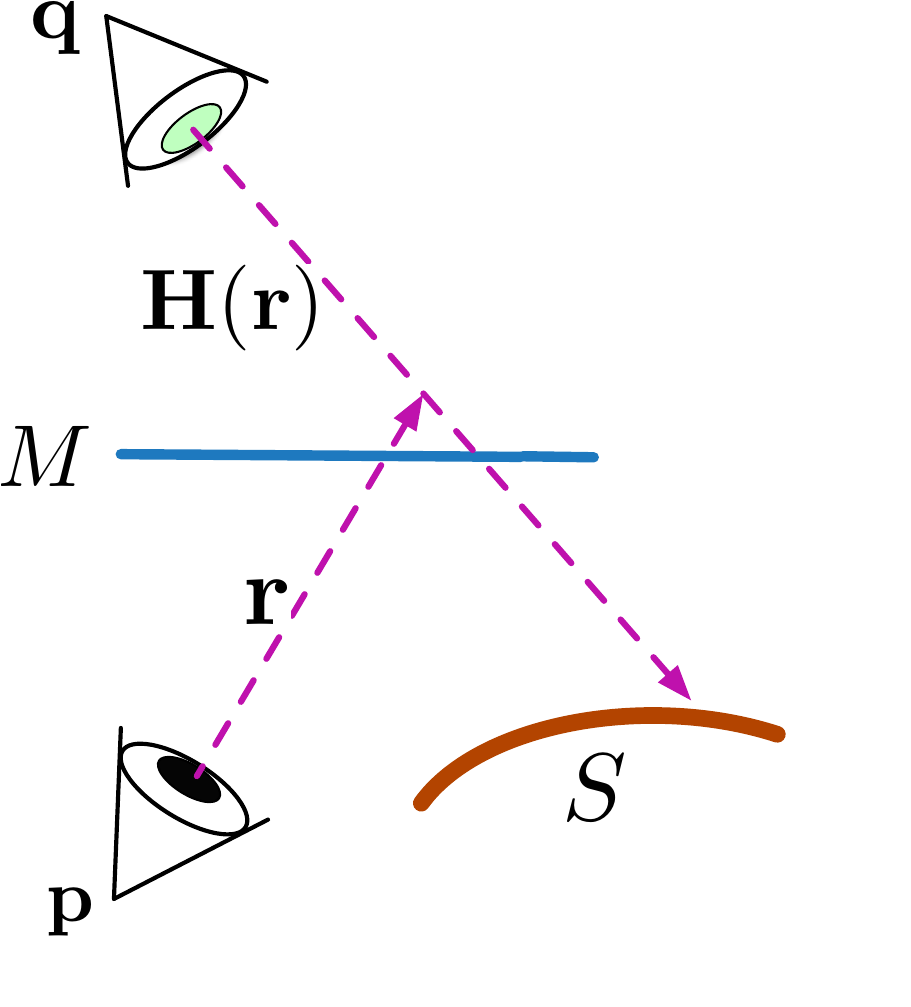}
{\scriptsize C.}\includegraphics[height=1.2in]{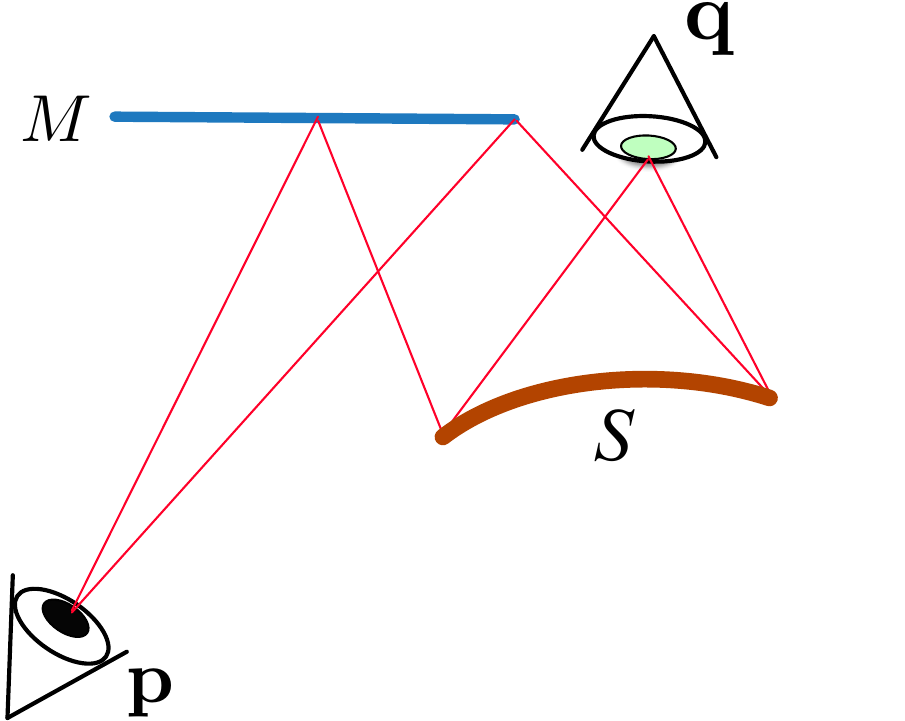}
\caption{A. Observer $\p$ views $S$ reflected in a mirror $M$, and $\q$ views $S$ directly (ignoring the obstruction by $M$). B. Ray $\ry$ corresponds to the ray $\hh{\ry}$. C. Here $\p$ and 
$\q$ have different fields of view, so ${\bf H}$ will need to contract the bundle leaving $\p$.}
\label{fig:virtual}
\end{figure}

\end{example}

\begin{example} \label{exam:eye}
\begin{figure}
\centering
\includegraphics[height=1.65in]{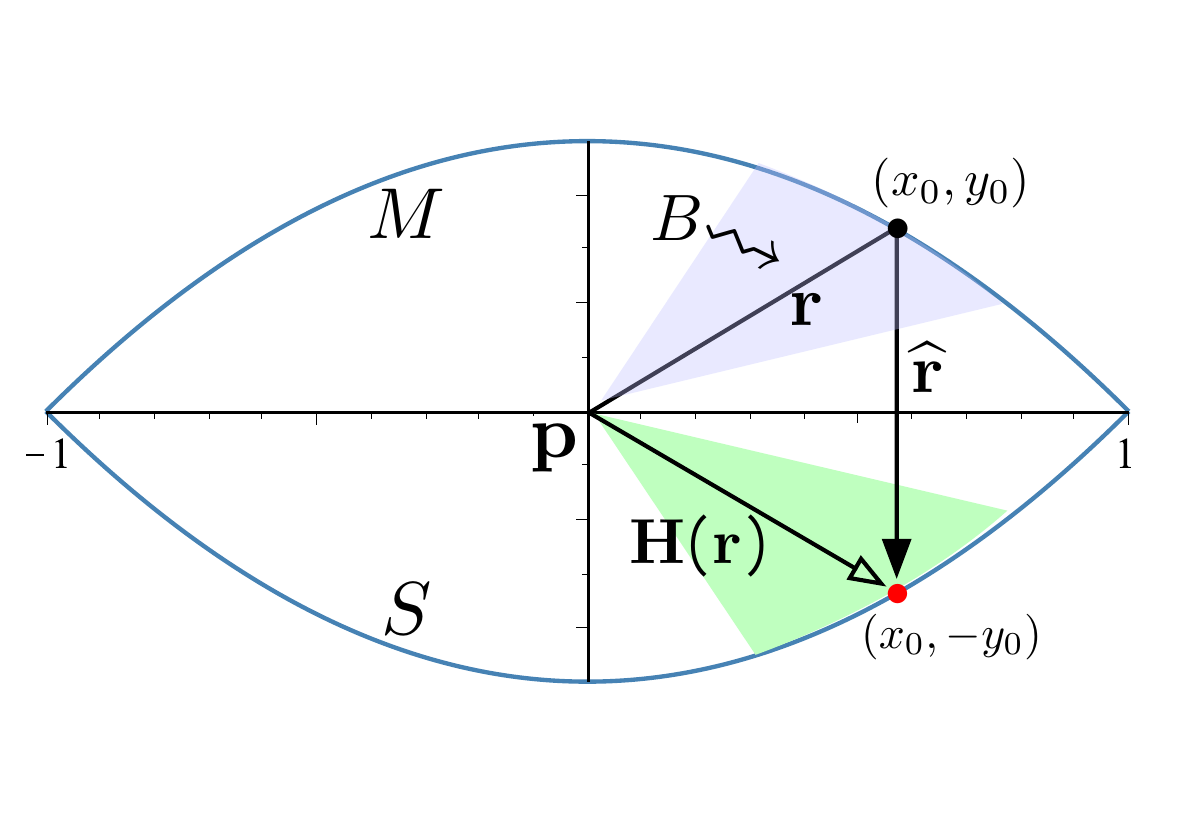}
\includegraphics[height=1.5in]{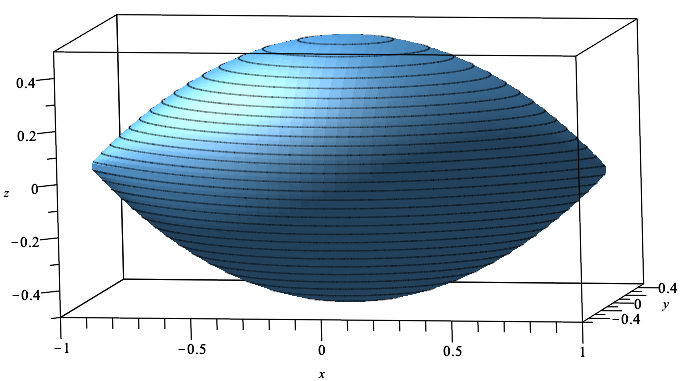}
\caption{On the left, an observer looking  along a ray ${\bf H}(\ry)$ sees a  red point on $S$. But, if the observer gazes up along $\ry$ at the mirror $M$,  they also see the red dot. On the right is plot of the eye, i.e., the surface swept out by revolving either of the parabolas about the horizontal axis. 
}
\label{fig:eye-example}
\end{figure}

Let's consider a problem with $\p=\q$. Take $S$ and $M$ be the parabolas given by $y=(x^2-1)/2$  and $y= (-x^2+1)/2$ respectively, on the interval  $(-1,1)$, as in the left of Fig. \ref{fig:eye-example}. Each has focus at $\p= (0,0)$, so a ray emanating from $(0,0)$ that strikes $M$ will be redirected downward by reflection.   Suppose that we are given a bundle of rays $B$ with common source $(0,0)$ that strike $M$.  An observer at $\p$ can look up along any of the rays in $B$.  Let $\ry$ be one such ray and $\bf H$ be the reflection
\[
{\bf H}(x,y) = (x,-y),
\]  
which flips the entire bundle down, so that $\p$ can also look down along the rays of $\hh{B}$ (which is colored light green). Thus $\p$'s view through $B$ and $\hh{B}$ are ``the same''. 

For example, in Fig. \ref{fig:eye-example}A, when the observer gazes down along $\hh{\ry}$ they see a {red dot}. But when they gaze up along $\ry$, they also see a {red dot}, since  the reflection of $\ry$ off of $M$, which is a vertical ray that we denote as \df{$\widehat{\ry}$}, intersects with $\hh{\ry}$ at the red dot. This is the key property of all eigenmirror/surface pairs $(M,S)$. Let's refer to a point $\x \in M$ as \df{physical} if it has this property, i.e., a ray $\ry$ from $\p$ that strikes $\x$ will reflect and intersect a positive multiple of ${\bf H}$. 

To extend this example to $\br^3$, we rotate $y=(x^2-1)/2$  about the $x$-axis, obtaining a surface of revolution which we refer to as the \df{eye} - see the right of Fig. \ref{fig:eye-example}. Note that the rotation fixes the focus at $(0,0,0)$. Reusing names, let $M$ be the upper half of $P$, and let $S$ be the lower half, and take 
\[
\hh{x,y,z} = (x,-y,-z),
\]
Then $P$ can be represented implicitly as
\begin{equation}
\sqrt {{y}^{2}+{z}^{2}}+\sqrt {{x}^{2}+{y}^{2}+{z}^{2}}=1.
\label{eqn:eye}
\end{equation} 
 
Raytracing simulations of a portion of $M$ and $S$ appear in Fig. \ref{fig:eye-simulations}. Here one can see what an observer at $\p$ inside the eye would see.
\begin{figure}
\centering
\includegraphics[height=1.95in]{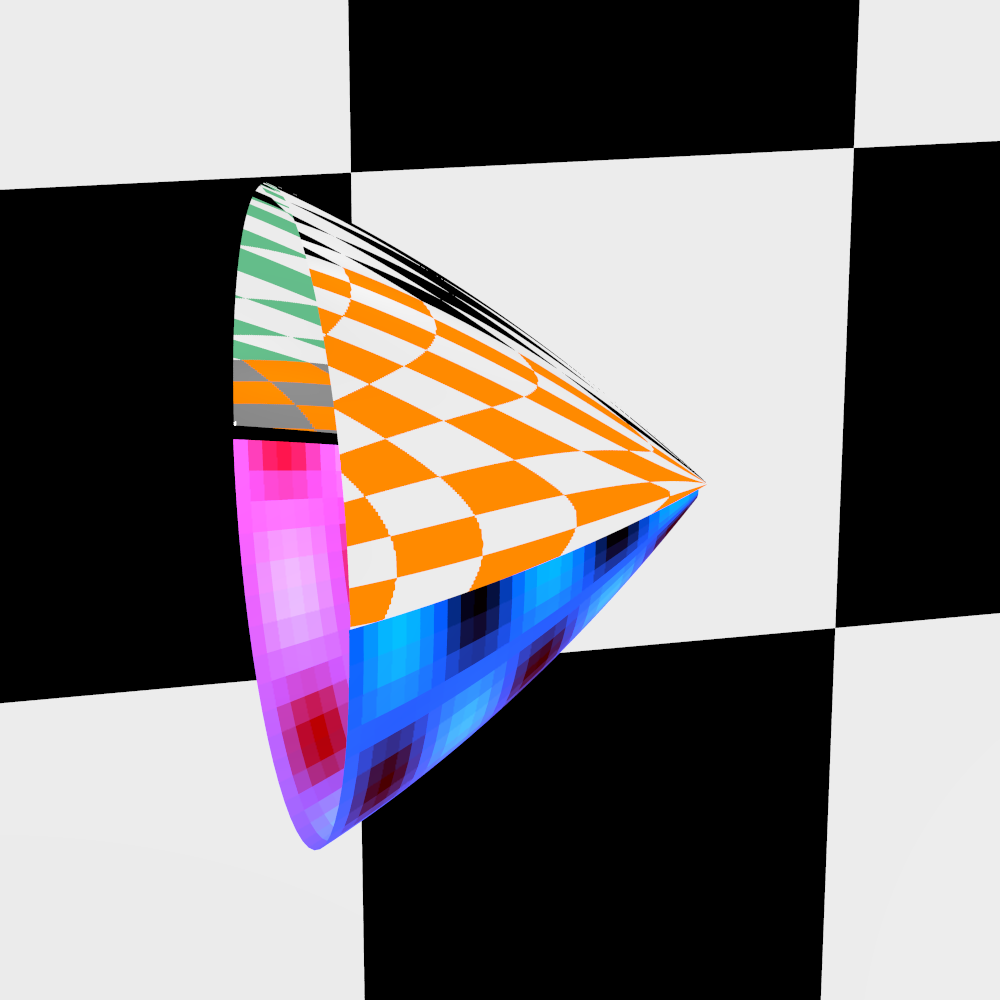}
\includegraphics[height=1.95in]{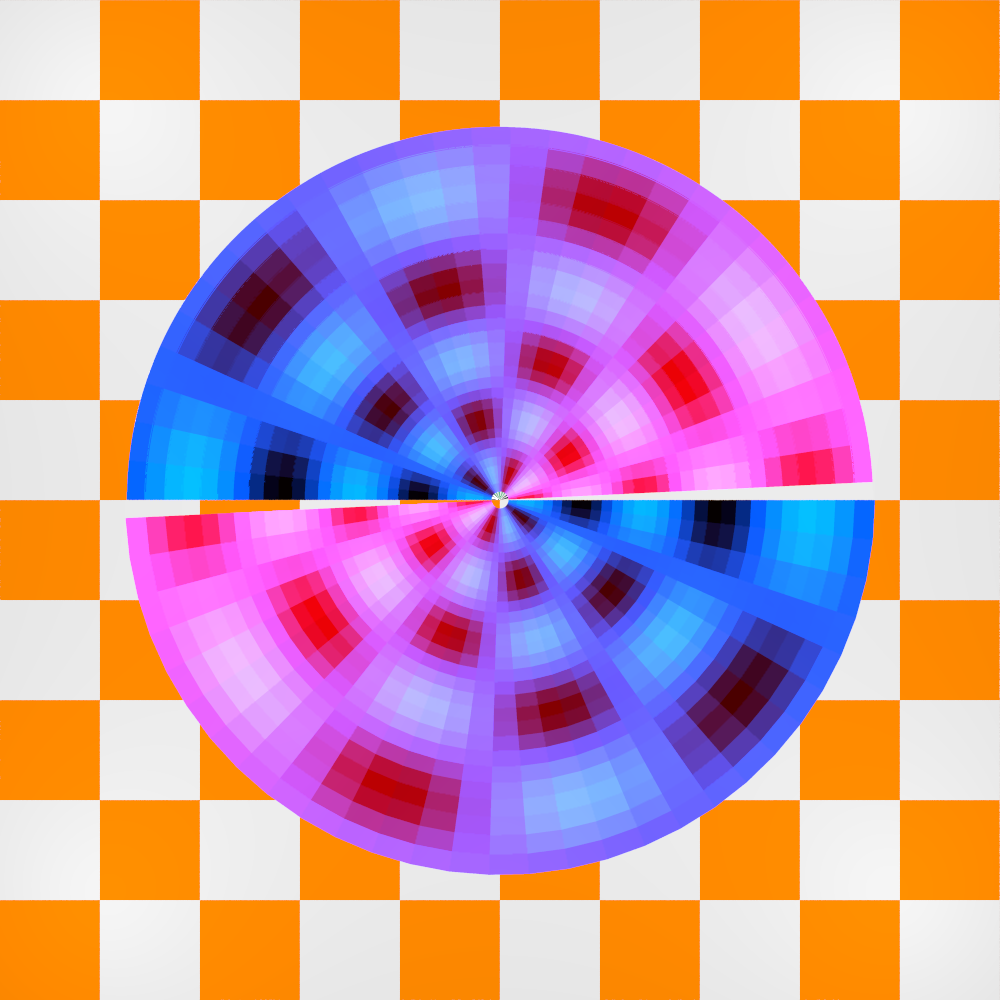}
\caption{Two raytracing simulations (POV-Ray) of a portion of $M$ and $S$ in a cubical test room with checkerboard walls. On the left, we see the surfaces from the side. The upper surface is $M$ and the lower surface $S$ has a pink and blue texture. On the right the observer views $M$ and $S$ from the origin, which demonstrates that $S$ is in fact an eigensurface of $M$.  Above $S$, the mirror $M$ appears to be identical to $S$, except for the action of the orthogonal map $\bf H$. 
}
\label{fig:eye-simulations}
\end{figure}

\label{exam:eye-exam}
\end{example}

{\bf Remarks.}
\begin{enumerate}

\item In example 1, we have $\p \neq \q$ and in example 2, $\p=\q$.
In this paper we will mostly consider the case of  $\p=\q$. The case of $\p \neq \q$ is  considered in \cite{hicks20josaa}.

\item We require that the {\em rays} $\hh{\ry}$ and $\widehat{\ry}$ intersect, although of course the calculations we will perform below involve intersecting the {\em lines} that contain these rays. Thus we run the risk of getting \df{mock solutions}\footnote{We call them mock solutions instead of false solutions because we will see that they are not entirely nonsense.} if we are not careful - see Fig. \ref{fig:eye-example-problem}A for an example. One may be tempted to remove the radicals from \ref{eqn:eye}, to obtain a polynomial representation: 
$
\left( {x}^{2}-1 \right) ^{2}=4\,{y}^{2}+4\,{z}^{2},
$
In the process, mock solutions are introduced - see Fig. \ref{fig:eye-example-problem}B. 
\begin{figure}
\centering
A. \includegraphics[height=1.05in]{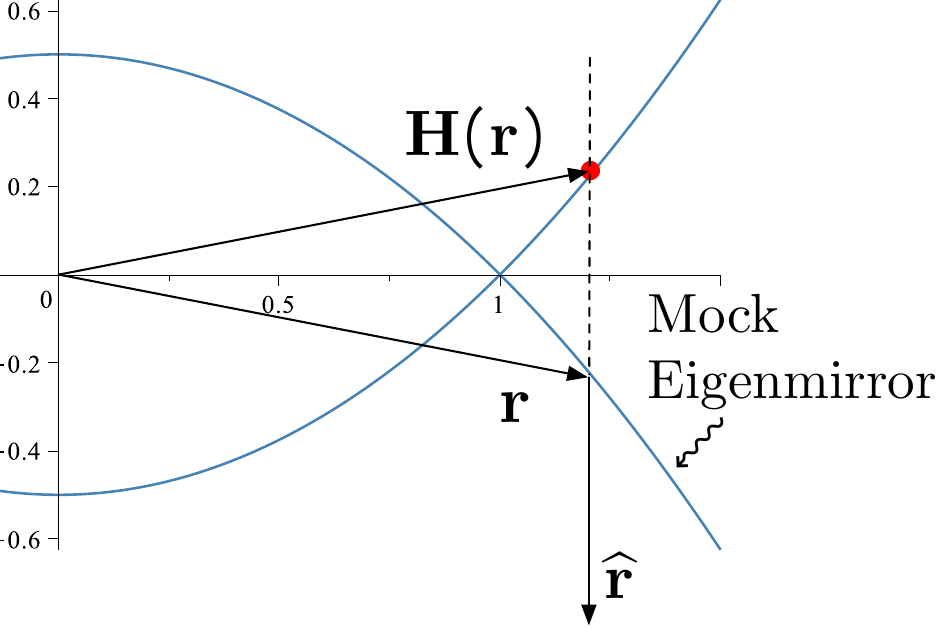}
B. \includegraphics[height=1.05in]{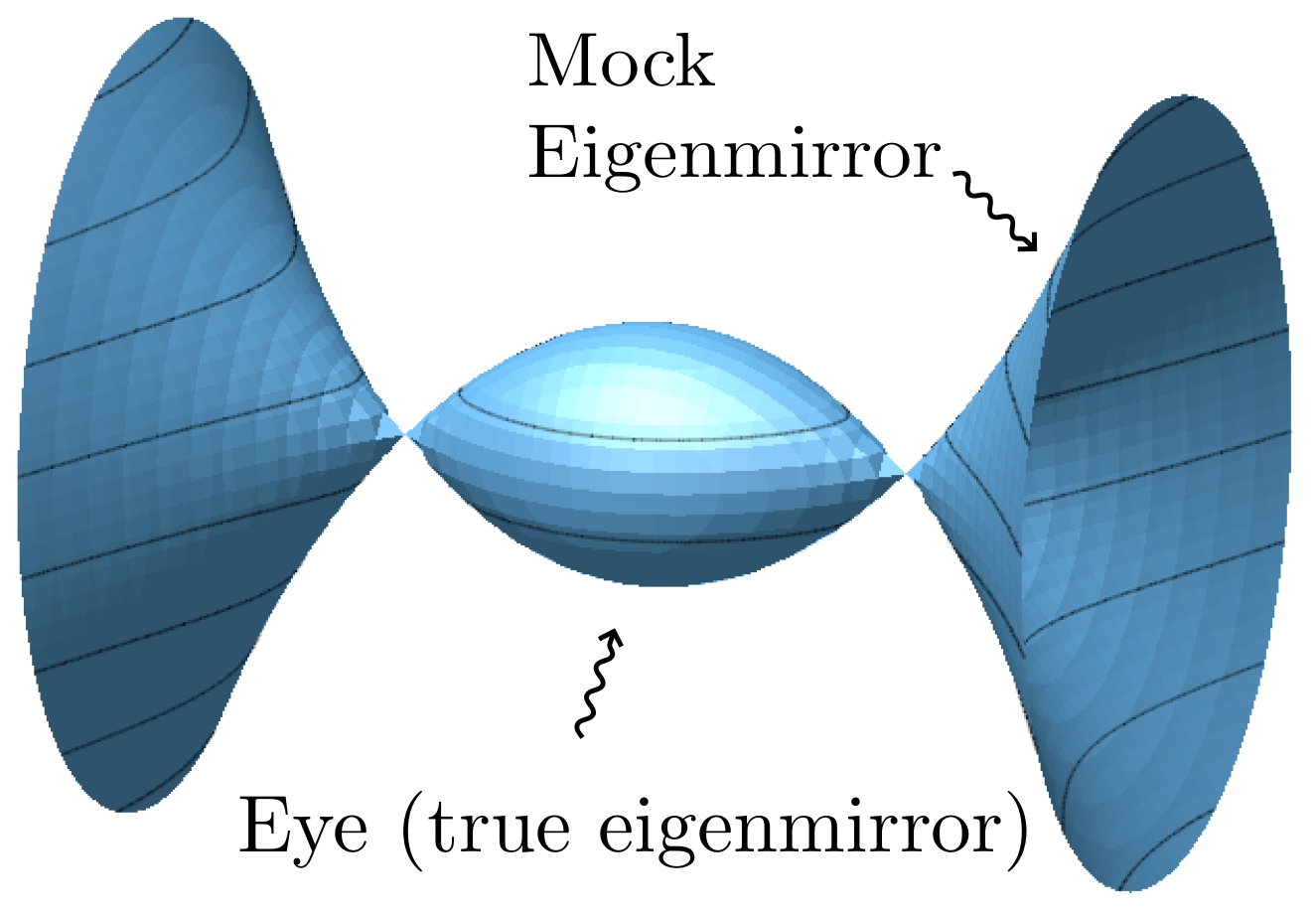}
C.\includegraphics[height=1.05in]{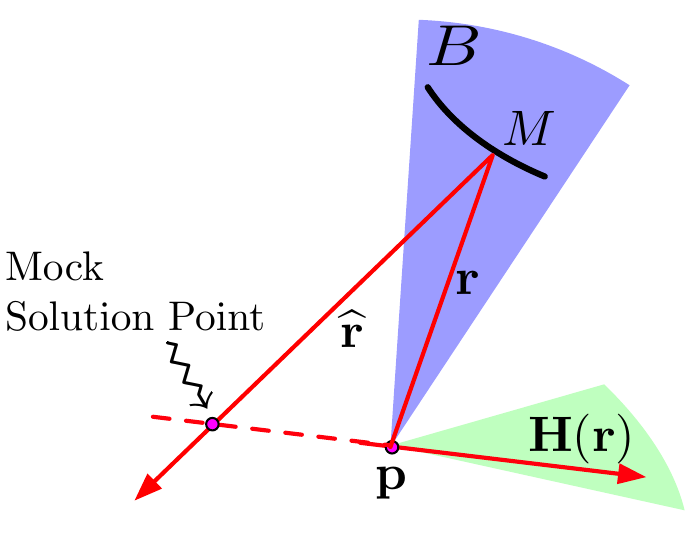}
\caption{A. If one extends the domain of the parabolas to $\br$, and the  red dot is moved on $y=(x^2-1)/2$ to have $x$ coordinate $>1$, then $\hh{\ry} \cap \widehat{\ry} = \emptyset$. In this diagram, the $\hh{\ry} \cap \widehat{\ry} = \emptyset$, i.e., physicality fails. B. The full surface of revolution. The eye in the center is a true solution to the problem, but the ``wings'' attached to it are mock solutions. C. An abstract depiction of how mock solutions can occur.
}
\label{fig:eye-example-problem}
\end{figure}

\item This example is extremely special, since $M$ and $S$ are geometrically congruent. In fact, we could take the $M$ and $S$ to be all sorts of subsets of the eye.

\item We do not distinguish between the ``front'' or ``back'' of $S$. One can think of $S$ as a thin fabric with the same texture on both sides.

\item In $\br^3$, if we are {\em given} $\bf H$ and an asymmetric $M$,  then  it unlikely that  that  $\hh{\ry}$ intersects $\widehat{\ry}$ - see Fig. \ref{fig:eigenmirror0-layout}. Therefore, we turn our attention to the problem of finding all pairs $(M,S)$ for a given ${\bf H}$.

\begin{figure}[htbp]
\centering
\includegraphics[height=1.5in]{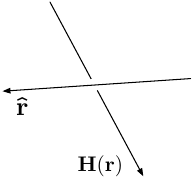}
\caption{In $\br^3$, given $M$ and $\bf H$, if $M$ has no symmetry then the probability  that $\widehat{\ry}$ and $\hh{\ry}$ intersect is zero.}
\label{fig:eigenmirror0-layout}
\end{figure}

\end{enumerate}

\section{Some Definitions and Notations}
\label{sec:definitions}
Here, we will err on the side of giving some definitions that might be familiar to the reader. 

\begin{define} 
If $d \in \bz$ and $f: \mathscr{W} \ra \mathscr{U}$ is a map between vector spaces with the property that

\begin{equation}
f(\lambda \x) = \lambda^d f(\x)
\label{eqn:homogeneous}
\end{equation}
for all $\x \in \mathscr{W}$ and $\lambda \neq 0$, then $f$ is said to be  \df{homogeneous of degree $d$}. 
\end{define}

\begin{remarks}\

\begin{enumerate} 
\item  $f_1,...,f_m:\br^n \ra \br$ are homogeneous of degree $d$ iff the map $\br^n \ra \br^m$, $\x \mapsto (f_1(\x),...,f_m(\x))$,   is homogeneous of degree $d$. 

\item  We may apply the adjective ``homogeneous'' when  $f$ is defined on a subset of $\mathscr{W}$ and $f$ satisfies \eqref{eqn:homogeneous}.
 
\end{enumerate}

\end{remarks}

\begin{define}
For $\q, \vv \in \br^3$,  we define a ray with \df{source} $\q$ and direction ${\bf v}$ to be 
$\ray{\q}{\vv} = \set{\q + \lambda \vv }{\lambda \geq 0}$. 
\end{define}
Of course $\ray{\q}{\vv} = \ray{\q}{\lambda\vv}$ for $\lambda>0$. 

{\em In diagrams, the ray $\ray{\q}{\vv}$ will be depicted as a point labeled $\q$ and an arrow coming out of it labelled $\vv$.}

Regarding optics, we need only the \df{law of reflection}, which says that reflection of ${\bf v}\in \br^n$ about  $\nn \in \br^n-\{{\z}\}$ is 
\begin{equation}
\reflect{{\bf v}, \nn} = -{\bf v} + 2 \frac{{\bf v}\cdot\nn}{\nn\cdot\nn} \nn.
\label{eqn:lawofreflection}
\end{equation}
Some useful facts are
\begin{enumerate}[a.]
\item $\reflect{{\bf -v}, {\nn}}=-\reflect{{\bf v}, {\nn}}$,
\item  
$
\reflect{{\bf v}, \lambda \nn} = \reflect{{\bf v}, \nn}
$, 
$\lambda \neq 0$,
\item If  $\norm{\bf v}  = \norm{\bf w} >0$, then  
\begin{equation}
\reflect{{\bf v}, {\bf v}+{\bf w}} = {\bf w} \text{ \quad and \quad  }
\reflect{{\bf w}, {\bf v}+{\bf w}} = {\bf v}. 
\label{eqn:reflectvaboutv+w=w}
\end{equation}
Equation \eqref{eqn:reflectvaboutv+w=w} expresses the fact that the reflection of light is  {\bf reversible}\index{reversibility of geometric optics}.
\end{enumerate}

\begin{figure}[htbp] 
   \centering \includegraphics[height=1.0in]{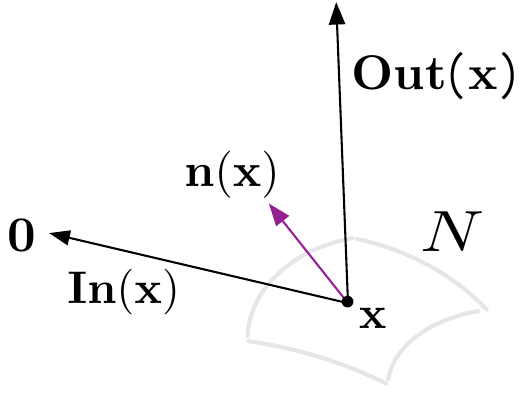} 
   \caption{Given a 2-manifold $N$, $\In{\x}$  may be reflected about $\bf n(x)$ to define $\Out{\x}$.}
   \label{fig:vectors-and-rays}
\end{figure}

\begin{define} Let  $N\subseteq \br^3-\{\z\}$ be a differentiable 2-manifold, $\x \in N$.
\begin{enumerate}[a.]
\item  We write \df{$\n({\x})$} for a normal to $N$ at ${\x}$, i.e., ${\n}$ is a vector field on $N$. (We don't require continuity.)
\item \df{$\In{\x}$} $= -\x$.

\item $\ssp{\mathbf{Out}}{N}({\x})  =  \reflect{\In{\x}, \n(\x)}$. \index{${\bf Out}_N(\x)$}

\end{enumerate} 
\end{define}
Thus, $\n({\x})$, $\In{\x}$, and $\mathbf{Out}_N({\x})$ are vectors, while, for example, $[\x, \mathbf{Out}_N({\x})]$ is the ray that reflects off of $N$ - see Fig. \ref{fig:vectors-and-rays}. Mostly, the manifold $N$ will be clear from context, so we will write $\Out{\x}$. Technically, $\n({\x})$ and $\mathbf{Out}_N({\x})$ only make sense in the presence of a given $N$.  $\In{\x}$ is independent of $N$.
\medskip

\section{Problem Statement} 

\begin{define}
Given a homogeneous map ${\bf H}:\br^3 \ra \br^3$, we define 
\[
\mathscr{E} = \mathscr{E}_{\bf H} = \{v \in \br^3 \mid (\exists \lambda \in \br)({\bf  H}(v)=\lambda v) \},
\] 
\end{define}

\begin{define}
Given a homogeneous map ${\bf H}:\br^3 \ra \br^3$ and a 2-manifold $N \subseteq \br^3 - \mathscr{E}$,   $\xz \in N$ is a  \df{physical point} of $N$ if  $[\z, \hh{\xz}]$ and   $[\xz, \Out{\xz}]$ intersect in a single non-zero point.
\end{define}

\begin{remarks}\

\begin{enumerate}
\item It does not make sense to discuss  physical points in the absence of a differentiable 2-manifold, since one need a normal to reflect light, i.e., $\Out{\xz}$ is  undefined without $N$.
\item Of course no points of $\mathscr{E}$  are ever physical for any $N$ since $N \subseteq \br^3 - \mathscr{E}$.
\end{enumerate}
\end{remarks}

\begin{define}
A single observer \df{eigenmirror problem} is a  homogeneous map ${\bf H}:\br^3 \ra \br^3$. A solution to an eigenmirror problem, also known as an \df{eigenmirror} of ${\bf H}$,  is  a differentiable 2-manifold $M\sseq \br^3-\mathscr{E}$, all of whose points are physical.
\label{def:solution} 
\end{define}
We will write
\[
\hh{x,y,z} = (h_1(x,y,z), h_2(x,y,z), h_3(x,y,z)).
\]
The observer, is, of course, at $\z$.

While $M$ is not a manifold with boundary, it may have boundary points, e.g.,  $M$ could be diffeomorphic to an open disk.

The requirement that $[\z, \hh{\x}] \cap [\x, \Out{\x}]$ be a single point is what we referred to as the {\bf  physicality} of $\x$ when discussing the red dot in example \ref{exam:eye}.

\begin{figure}[htbp] 
   \centering \includegraphics[height=1.3in]{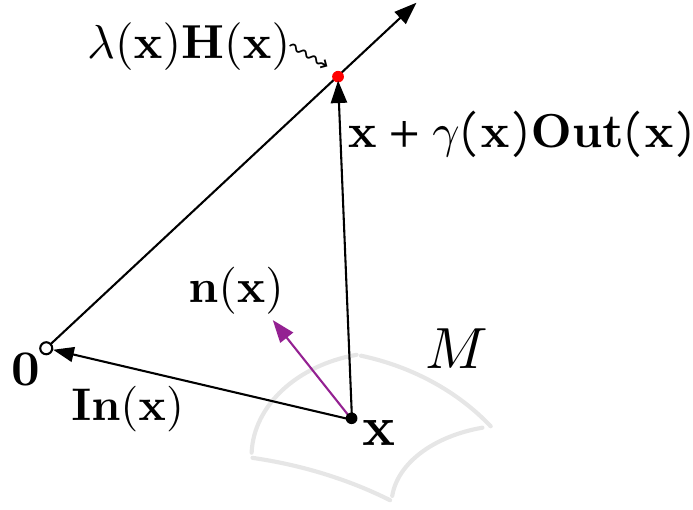} 
   \caption{$M$ is a solution to our problem,  if for each $\x \in M$, there are $\lambda(\x)$, $\gamma(\x)>0$ such that $\lambda(\x) \hh{\x} = \x + \gamma(\x) \Out{\x}$. Thus the vectors $\In{\x}, \hh{\x}, \Out{\x}, {\bf n}(\x)$ all lie in the same two dimensional subspace.}
   \label{fig:lambdaHgammaOut}
\end{figure}

If $M$ is an eigenmirror of ${{\bf H} }$, then for $\x\in M$ 
\begin{equation}
\lambda(\x) \hh{\x} = \x + \gamma(\x) \Out{\x},
\label{eqn:intersectingrays}
\end{equation}
for one and only one pair of numbers $\lambda(\x),\gamma(\x) > 0$ - see Fig. \ref{fig:lambdaHgammaOut}. These are the side inequalities mentioned in the introduction.

It's worth putting down a short summary of how our vector fields are related.
\begin{lemma} 
\label{lem:HInOutn}
Suppose ${{\bf H} }$ is given, along with an eigenmirror $M \subseteq \br^3 -  \mathscr{E}$. 
Then for  $\x \in M$, 
${\bf n}(\x)$ and $\Out{\x}$ lie in $ \text{span}\{\In{\x} ,\hh{\x} \}$.
\end{lemma}
{\bf Proof.}
We have that
\begin{align}
\Out{\x} & = \reflect{\In{\x}, {\bf n}(\x)}\\
                  & = -\In{\x} + 2 \frac{\In{\x}\cdot {\bf n}(\x)}{{\bf n}(\x) \cdot {\bf n}(\x)} {\bf n}(\x).
\label{eqn:reflectInnablaphi}
\end{align}
We claim that $\In{\x} \cdot {\bf n}(\x) \neq 0$, because otherwise $\Out{\x} = -\In{\x}$, which is impossible since $\x \not \in \mathscr{E}$. 
Thus 
\[
{\bf n}(\x)\in \text{span}\{\In{\x} ,\Out{\x}\}.
\]

Additionally,  $\lambda(\x)\hh{\x} = \x + \gamma(\x) \Out{\x} = -\In{\x} + \gamma(\x) \Out{\x}$, so 

\[
\Out{\x} \in \text{span}\{\In{\x} ,\hh{\x} \}
\] 
and
\[
{\bf n}(\x) \in \text{span}\{\In{\x} ,\hh{\x} \}.
\] 
\QED

Notice that if $\xz \in M$, ${\bf n}(\xz)=\tau \xz$, $\tau >0$, $\xz  \neq \z$, then
\[
\Out{\xz} =
\reflect{-\xz, \xz} = 
 \xz + 2 (-\xz \cdot \xz) /(\xz \cdot \xz) \xz = 
-\xz. 
\]
Therefore, $[\xz , \Out{\xz}]=[\xz , -\xz]$ intersects $[\z, \hh{\xz}]$ at $\z$. Thus it can never be that  ${\bf n}(\xz)=\tau \xz$ for $\xz$ on an eigenmirror. In other words, if $\z$ is a source light, then an eigenmirror can never reflect light directly back at $\z$

Writing $\Out{\x} = (\theta_1(\x), \theta_2(\x), \theta_3(\x))$,   if \eqref{eqn:intersectingrays} holds, then we can solve for $ \gamma(\x), \lambda(\x)$:
\begin{equation}
 \gamma(\x)=-{\frac {\theta_1(\x)y-\theta_2(\x)x}{\h_1(\x)\theta_2(\x)-\h_2(\x)\theta_1(\x)}}, 
\label{eqn:gamma>0}
\end{equation}
\begin{equation}
\lambda(\x)=-{\frac {\h_1(\x)y-\h_2(\x)x}{\h_1(\x)\theta_2(\x)-\h_2(\x)\theta_1(\x)}}.
\label{eqn:lambda>0}
\end{equation}
Thus, if $M$ is known, $\bf Out$ is known,  and $\gamma$ and  $\lambda$ are given by \eqref{eqn:lambda>0} and \eqref{eqn:gamma>0}.
\begin{lemma}
\label{lem:triangle}
If $\x$ lies on an eigenmirror of ${\bf H}$, then
\[
|\In{\x}| +\gamma(\x) |\In{\x}| > |\hh{\x}| \lambda(\x).
\]
\end{lemma}

{\bf Proof.} This is just the triangle inequality for the triangle of Fig. \ref{fig:lambdaHgammaOut}, along with the facts that $|\Out{\x}| =  |\In{\x}|$, and that the triangle cannot be degenerate since $\x \in \br^3-\mathscr{E}$. \QED
\begin{define}
Corresponding to the eigenmirror $M$, we define the  \df{eigensurface} $S$ to be the \df{parametrized surface} given by
\[
\x \mapsto  \lambda({\x}) \hh{\x},\ \x \in M.
\]
\end{define}
Note that if we had not excluded $\mathscr{E}$ from our problem domain, we could have $\hh{\xz} = -\xz$, which would result in   $|[\z, \hh{\xz}] \cap [\xz, \Out{\xz}]|$ being infinite, and so making the definition of $S$ ambiguous. We will see below other reasons for putting $\mathscr{E}$ aside.

A natural concern at this point is that  $M$ and $S$ may obstruct rays. We  ignore this issue, but the reader may take some solace, though, in the fact that if $M$ is an eigenmirror of  ${{\bf H} }$, and $N \subset M$ is a submanifold of $M$, then  $N$ is an eigenmirror of  ${{\bf H} }$. Thus  one is free to shrink $M$, possibly obtaining a surface that is close to a flat disk.

\section{The Anti-Eikonal Equation}
Given ${\bf H}$ and $M$,  $\x \in M$, we know that ${\bf n}(\x)$  lies in $ \text{span}\{\In{\x} ,\hh{\x} \}$. Thus,
\begin{equation}
(\In{\x} \times \hh{\x})\cdot {\bf n}(\x) = 0.
\label{eqn:preaee}
\end{equation}
What this equation is telling us, is whether two parametric lines 
$t \ra t\hh{\x}$ and $s \ra \x + s \Out{\x}$ intersect.

\begin{figure}[htbp] 
   \centering \includegraphics[height=1.6in]{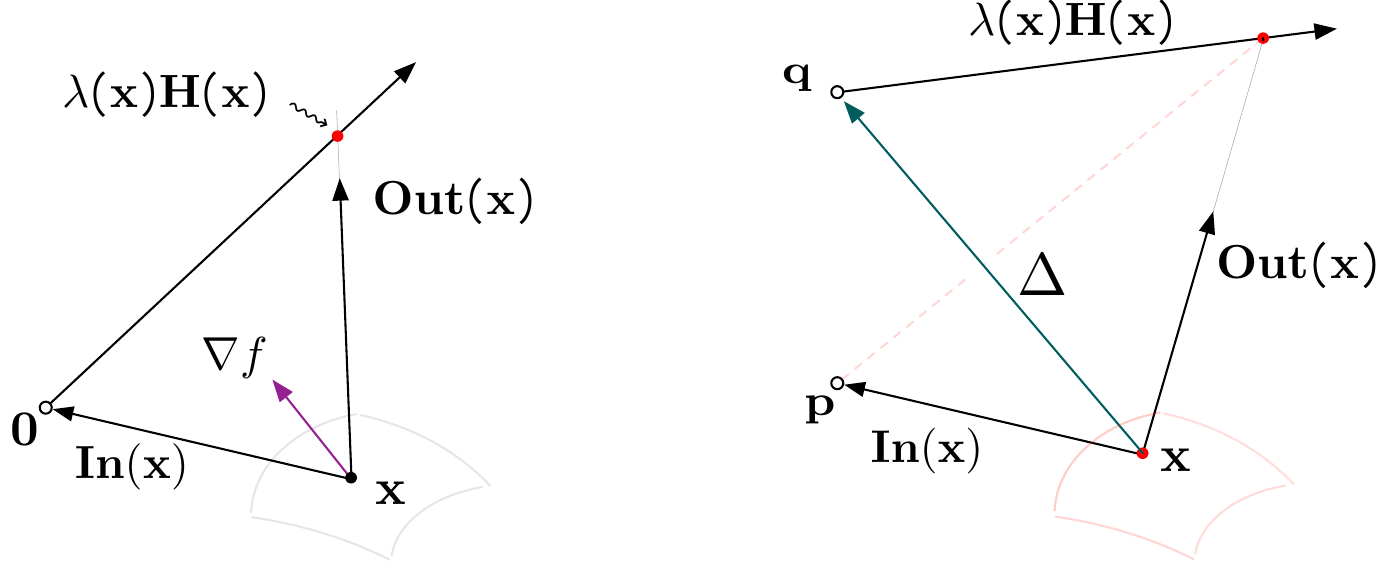}
   \caption{On the left, the derivation of the one-point anti-eikonal equation, and on the right the two-point equation.}
   \label{fig:aee}
\end{figure}

Now suppose that we are given a differentiable $\phi: \br^3-\mathscr{E} \ra \br$, and 
\[
M \subseteq \{\x\in \br^3-\mathscr{E} \mid \phi(\x)=C\}.
\]
 Then for $\x \in M$ we have
\begin{equation}
\left[\In{\x}\times \hh{\x}\right]\cdot \nabla \phi(\x) = 0. \tag{AEE}
\label{eqn:aee}
\end{equation}
This is the {\bf anti-eikonal equation  (AEE)} - see  Fig. \ref{fig:aee}.

On the other hand, if we are given ${\bf H}$, but not $M$, and treat $\phi$ in the \ref{eqn:aee} as  unknown, then the \ref{eqn:aee} is a linear PDE for $\phi$.\footnote{If one prefers the eigenmirror to be given as the graph of a function, $u(x,y)$,  then the \ref{eqn:aee} becomes quasilinear, as the  $\x$ in in $\In{\x}\times \hh{\x}$ is now $(x,y, u(x,y))$.}
In coordinates, the \ref{eqn:aee} is

\begin{equation}
\left(z\h_2- y\h_{3}  \right) \phi_x + 
\left(x \h_{3}-z\h_{1} \right) \phi_{{y}} + (y\h_1-x\h_2)\phi_z  = 0. 
\label{eqn:aee1}
\end{equation}

{\em However, solving the \ref{eqn:aee} does not necessarily yield an eigenmirror of ${\bf H}$, since the \ref{eqn:aee} only detects if lines intersect, not rays.} 

Thus a solution of the \ref{eqn:aee}  may be a mock solution, like the degree four polynomial $
\left( {x}^{2}-1 \right) ^{2}=4\,{y}^{2}+4\,{z}^{2},
$ that includes both the eye and the wings, from example \ref{exam:eye}.  That discussion led to the  requirement that $\lambda(\x),\gamma(\x) > 0$. Now that we have proper definitions though, we can see that a mock solution is simply a solution to the \ref{eqn:aee} that is not an eigenmirror of ${\bf H}$. 


Observe that if $\phi$ is a solution to the \ref{eqn:aee}, then so is $\phi-C$, $C\in \br$. Also, $(x_0,y_0,z_0) \in \Phi_{\phi(x_0,y_0,z_0)}$.

The characteristic equations for the \ref{eqn:aee} are $\dot{\x} = \In{\x}\times \hh{\x}$, i.e.,
\begin{align} 
\begin{split}
\dot{{x}} & = z\h_2- y\h_{3},\\ 
\dot{{y}} & = x \h_{3}-z\h_{1},\\
\dot{{z}} & = y\h_1-x\h_2.
\end{split}
\label{eqn:characteristics}
\end{align}
For example, if ${\bf H}$ is linear then the above system is quadratic and homogeneous.

We define the \df{characteristic vector field of ${\bf H}$} to be
\[
\V(\x) = \In{\x}\times \hh{\x}  =  (z\h_2- y\h_{3}, x \h_{3}-z\h_{1}, y\h_1-x\h_2).
\]
$\V$ is homogeneous and $\V(\x)\cdot \x=0$. Thus  $R(x,y,z)=x^2+y^2+z^2$ is a first integral of $\dot{\x} = \In{\x}\times \hh{\x}$.\footnote{Just as the {\bf total kinetic energy}\index{kinetic energy of a rigid body} and the {\bf total angular momentum}\index{angular momentum of a rigid body} are for the free rigid body.} 
$\V$ is of course naturally defined on all of $\br^3$, and $\V$ never vanishes on $\br^3 - \mathscr{E}$.

$\V$ is  tangent to every sphere $S^2(r)$, $r>0$, but also to all the eigenmirrors of ${\bf H}$. In fact, if one  is careful about choosing a curve of initial conditions for $\V$, examples can be generated by flowing off of that curve in the direction of $\V$  - see  \cite{hicks20josaa}.

\begin{remark}
Here we only consider the one-point problem, with the observer at $\z$.  It is possible, as in Fig. \ref{fig:cylinder},  to formulate a {\bf two-point problem}. We include here the {\bf two-point anti-eikonal equation} for comparison. The derivation  is similar to the derivation of the \ref{eqn:aee}, except that we can't simply take a dot product with $\nabla \phi$ - instead it must be with $\Out{\x}$, which leads to the equation being nonlinear:
\begin{equation}
(\Delta \times \hh{\x}) \cdot \Out{\x} = 0.\tag{AEE2}
\label{eqn:aee2}
\end{equation}
See the right of Fig. \ref{fig:aee} for the geometric explanation. 
In coordinates, we may, with no loss of generality take $\p=\z$ and $\q=(0,0,1)$, so that \eqref{eqn:aee2} becomes
\begin{dmath}
\left( 2\,xyh_{{3}}-2\,xzh_{{2}}+xh_{{2}}+yh_{{1}} \right) {f_{{x}}}^
{2}+ \left( -2\,{x}^{2}h_{{3}}+2\,xzh_{{1}}+2\,{y}^{2}h_{{3}}-2\,yzh_{
{2}}-2\,xh_{{1}}+2\,yh_{{2}} \right) f_{{x}}f_{{y}}+ \left( 2\,{x}^{2}
h_{{2}}-2\,xyh_{{1}}+2\,yzh_{{3}}-2\,{z}^{2}h_{{2}}+2\,h_{{2}}z
 \right) f_{{x}}f_{{z}}+ \left( -2\,xyh_{{3}}+2\,yzh_{{1}}-xh_{{2}}-yh
_{{1}} \right) {f_{{y}}}^{2}+ \left( 2\,xyh_{{2}}-2\,xzh_{{3}}-2\,{y}^
{2}h_{{1}}+2\,{z}^{2}h_{{1}}-2\,h_{{1}}z \right) f_{{y}}f_{{z}}+
 \left( 2\,xzh_{{2}}-2\,yzh_{{1}}-xh_{{2}}+yh_{{1}} \right) {f_{{z}}}^
{2}=0.
\end{dmath}
\end{remark}

\section{Invariance, Vector Fields, and Physicality}

Recall, from equation \eqref{eqn:preaee}, that for a given ${{\bf H} }$ and eigenmirror $M$, for $\x \in M$,  ${\bf n}(\x)$ is  perpendicular to {\em both} $\In{\x}$ and $\hh{\x}$. Putting aside $M$, if one has a vector field $\om{\x}:\br^3 \ra \br^3$ of the form
\begin{equation}
\om{\x}  =  \omega_1(\x) \In{\x} + \omega_2(\x) \hh{\x}
\label{eqn:omegaInH}
\end{equation} 
with $\curlnp \Omega = \z$, $\omega_1:\br^3 \ra \br$ and $\omega_2:\br^3 \ra \br$ differentiable, then there exists  a potential $\phi:\br^3 \ra \br$ such that $\nabla \phi = \Omega$ \cite{cantarella02maa}. Here there is no topology to trip one up when seeking a global potential, even though when defining eigenmirrors we worked in $\br^3 - \mathscr{E}$.  For example, in computational experiments performed by the author,   $\omega_1(\x)$, $\omega_2(\x)$, and $\hh{\x}$ are usually taken to be polynomials, so $\Omega$ is a polynomial vector field and there is no issue with finding $\phi$ as long as $\curlnp \Omega = \z$, because we will end up restricting these quantities to $\br^3 - \mathscr{E}$.

In what follows, if we mention $({\bf H}, \Omega)$ then it should be assumed that
the above notations and definitions are in effect.

\begin{example} \label{exam:eye2}

Continuing with  example \ref{exam:eye}, we had the degree four singular surface
\[
\phi = \left( {x}^{2}-1 \right) ^{2} - 4{y}^{2} - 4{z}^{2} =0,
\]
with corresponding gradient 
\[
\nabla \phi = (4x \left( {x}^{2}-1 \right) ,-8y,-8z).
\]
Since  ${\bf H}(x,y,z) = (x,-y,-z)$ in this example, it must be that 
$\omega_1=-2{x}^{2}+6$, $\omega_2=2{x}^{2}+2$.
\end{example}

\begin{define} Given $({\bf H}, \Omega)$, let
\[
\mathscr{O}_2=\{\x \in \br^3 \mid \omega_2 (\x) =0 \}.
\]
\end{define}

\begin{define}
Given $({\bf H}, \Omega)$, let $\Phi_C = \{ \x \in \br^3 - (\mathscr{E} \cup \mathscr{O}_2) \mid \phi(\x)=0\}$. Then an {\bf eigenmirror} of $({\bf H}, \Omega)$ is an eigenmirror $M$ of ${\bf H}$ with $M \subseteq \Phi_C$ for some $C$.
\end{define}

Observe that both $\mathscr{E}$ and $\mathscr{O}_2$ are closed in $\br^3$.

\begin{lemma}
If $M$ is an eigenmirror of $({\bf H}, \Omega)$, then  $M \cap \mathscr{O}_2 = \emptyset$. 
\end{lemma}

{\bf Proof.}
If $\xz \in \mathscr{O}_2$, then $\om{\xz}$ is a multiple of $\In{\xz}$, so $\xz$ cannot be physical.
\QED
\medskip 

\begin{lemma}
$\Omega$ does not vanish on $\br^3 - (\mathscr{E} \cup \mathscr{O}_2)$.
\end{lemma}

{\bf Proof.} Suppose that for some $\xz \in \br^3 - (\mathscr{E} \cup \mathscr{O}_2)$ we had
\begin{equation}
\omega_1(\xz) \In{\xz}  = -\omega_2(\xz) \hh{\xz}.
\label{eqn:omega1In=-omega2H}
\end{equation}
Say both sides of \eqref{eqn:omega1In=-omega2H} are zero. Then $\omega_2(\xz)=0$ or $\hh{\xz}=0$. Both of those possibilities are forbidden since $\xz \in \br^3 - (\mathscr{E} \cup \mathscr{O}_2)$.

On the other hand, if both sides of \eqref{eqn:omega1In=-omega2H} are not zero then $\hh{\xz}$ is a non-zero multiple of $\xz$, which is impossible since $\xz \in \br^3 - (\mathscr{E} \cup \mathscr{O}_2)$. 
\QED
\medskip

\begin{example} (Continuation of example \ref{exam:eye2})
Recall that in example \ref{exam:eye2}, we saw that $\Omega = [4\,x \left( {x}^{2}-1 \right) ,-8\,y,-8\,z]$ and $\omega_1=-2\,{x}^{2}+6$, $\omega_2=2\,{x}^{2}+2$. 
Here $\mathscr{E}$ is the $x$-axis and $\mathscr{O}_2$ is empty.   On $\mathscr{E}$, $\Omega=(4\,x \left( {x}^{2}-1 \right) ,0,0)$, which vanishes at $(0,0,0), (-1,0,0)$ and $(1,0,0)$. 
So certainly $\Omega$ can vanish on $\mathscr{E}$.
\end{example}
\medskip

Since $\Omega$ does not vanish on $\br^3 - (\mathscr{E} \cup \mathscr{O}_2)$,  each ${\Phi_C} $ is a 2-manifold  \cite{docarmo}.
Also, every $\phi$ is a solution of the \ref{eqn:aee} since
\[
(\In{\x}\times \hh{\x})\cdot \nabla \phi = (\In{\x}\times \hh{\x})\cdot (\omega_1(\x) \In{\x} + \omega_2(\x) \hh{\x}) = 0.
\]

Therefore, every  ${\Phi_C}$ 
is a candidate for containing an eigenmirror.    

\begin{figure}[htbp] 
   \centering \includegraphics[height=3in]{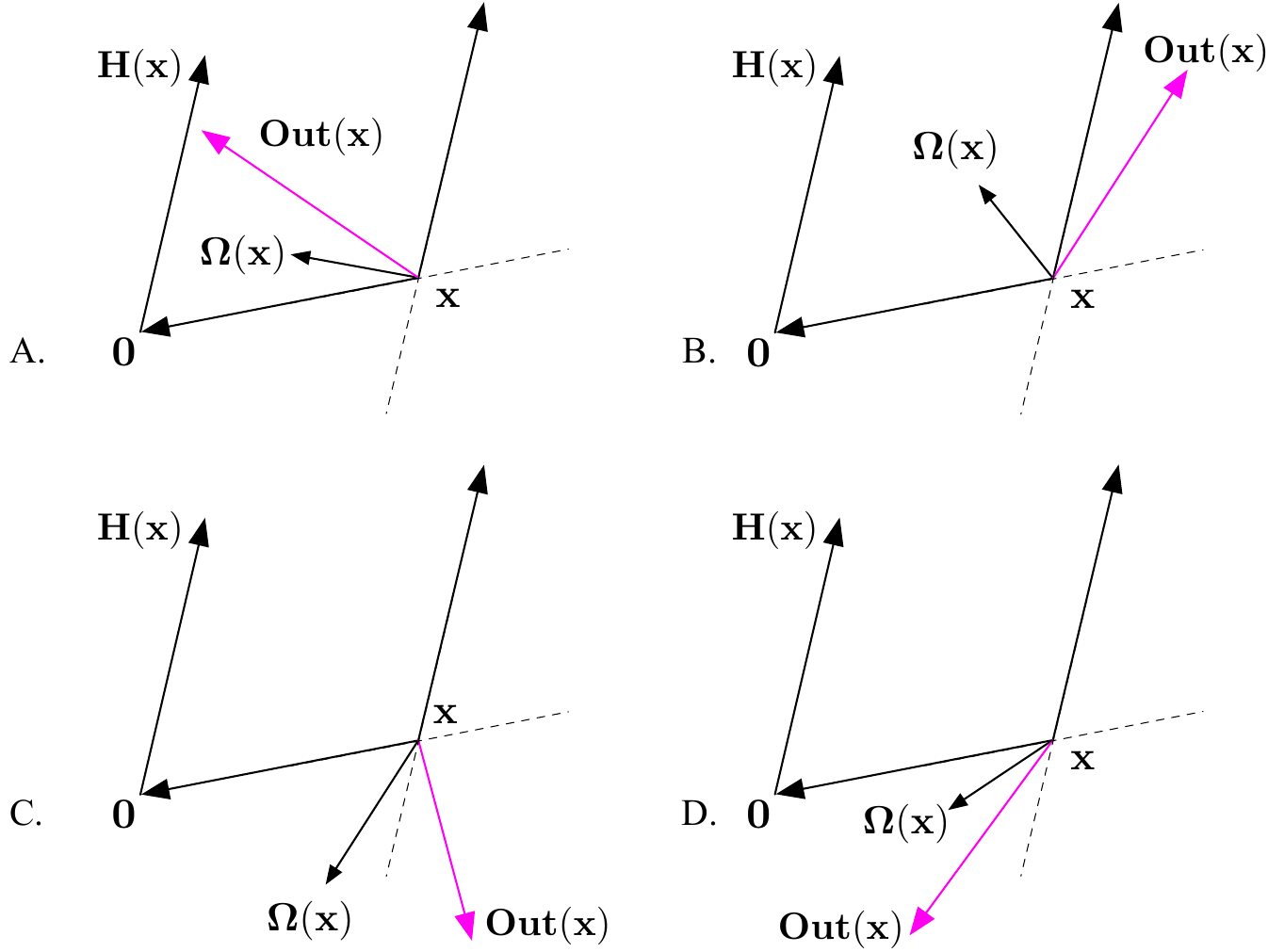} 
   \caption{A. For a fixed $\x$, the situation is planar. In (A)  $[\x, {\bf Out}(\x)]$ intersects $[\z, \hh{\x}]$, and so $\x$ is physical, but in the three other cases $\x$ is not physical.
   }
   \label{fig:In-Omega-H}
\end{figure}

For $\x \in {\Phi_C}$, since
$\{ \om{\x}, \Out{\x}, \In{\x}, \hh{\x}  \}$ all lie in the same 2D subspace and there is a reasonable chance that $[\x, {\bf Out}(\x)]$ will have non-zero intersection with 
$[\z, \hh{\x}]$. This means  the situation may be as in Fig. \ref{fig:In-Omega-H}A, as opposed to  \ref{fig:In-Omega-H}B, C, D. 


\begin{figure}[htbp] 
\centering \includegraphics[height=1.6in]{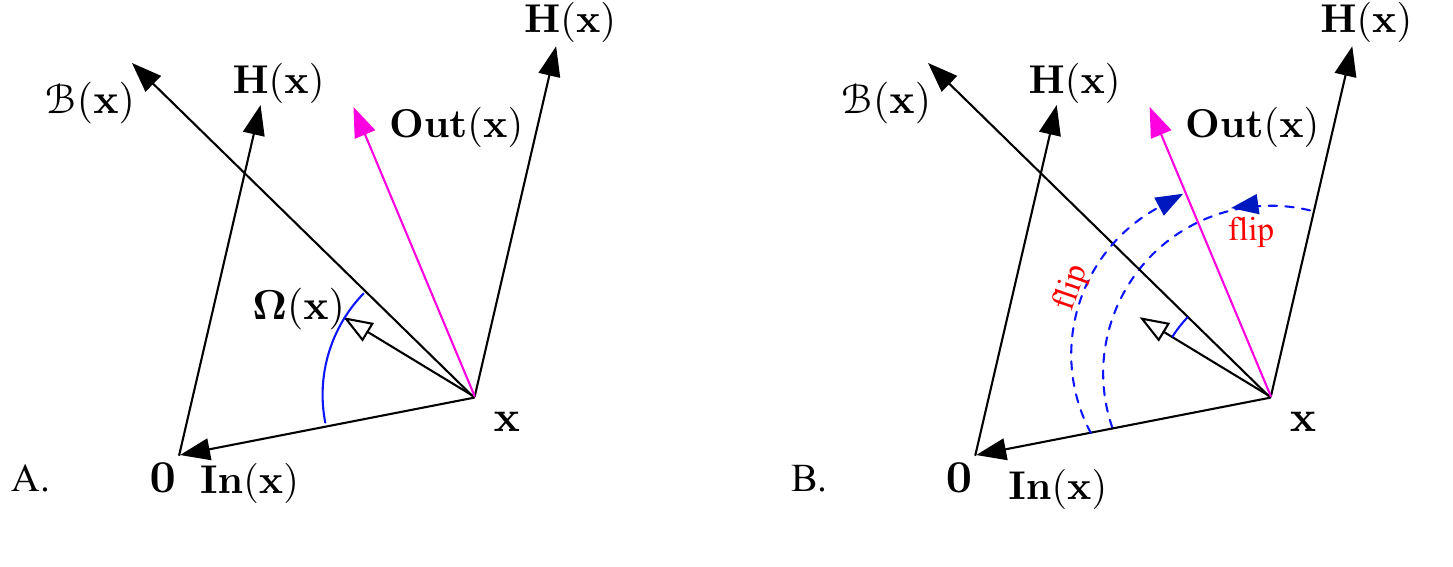} 
\caption{A. $\x$ will be a physical point as long as $\om{\x}$ lies between $\In{\x}$ and $\mathscr{B}(\x)$, the bisector at $\x$. B. The reflection of $\In{\x}$ about $\om{\x}$ lies in the open cone at $\x$ determined by $\In{x}$ and $\hh{\x}$. This follows from the ``two reflections is a rotation'' theorem.}
\label{fig:boundaryfield}
\end{figure}

\begin{define} For a given ${{\bf H} }$ we define the \df{bisector field} on $\br^3 - \mathscr{E}$ to be
\[
\mathscr{B}(\x) = \frac{\In{\x}}{\norm{\In{\x}}} + \frac{\hh{\x}}{\norm{\hh{\x}}}. 
\]
\end{define}

The idea here is that $\x$ is physical iff $\Omega(\x)$ lies in the interior of the cone with vertex at $\x$ that is determined by $\In{\x}$ and $\mathscr{B}(\x)$ -  see Fig. \ref{fig:boundaryfield}A. This might also be guessed by examining Fig. \ref{fig:In-Omega-H}.

\begin{lemma} 
For   $({\bf H}, \Omega)$,  suppose that for $\x \in \br^3 - (\mathscr{E} \cup \mathscr{O}_2)$, we  write
\[
\om{\x}= {\alpha_1}(\x) \In{\x} + {\alpha_2}(\x) \mathscr{B}(\x).
\]
Then  $\x \in \br^3 - (\mathscr{E} \cup \mathscr{O}_2)$  is a physical point of $\Phi_{C}$ if and only if 
\[
{\alpha_1}(\x){\alpha_2}(\x)>0.
\]  
\end{lemma}

{\bf Proof.} If ${\alpha_1}(\x){\alpha_2}(\x)>0$, then either ${\alpha_1}(\x)$, ${\alpha_2}(\x)>0$ or ${\alpha_1}(\x)$, ${\alpha_2}(\x)<0$. Let's say ${\alpha_1}(\x)$, ${\alpha_2}(\x)>0$.  

By \eqref{eqn:reflectvaboutv+w=w},  $\reflect{\hh{\x}, \mathscr{B}(\x)} = \In{\x}$. Therefore the reflection of $\In{\x}$ about $\om{\x}$ is the composition of a reflection of  $\hh{\x}$ about $\mathscr{B}(\x)$ followed by a reflection  about $\om{\x}$ - see Fig. \ref{fig:boundaryfield}B. Such a product of reflections is equal to a rotation by twice the angle between $\om{\x}$ and $\mathscr{B}(\x)$. But the angle between $\om{\x}$ and $\mathscr{B}(\x)$ is less than one-half the angle between $\In{\x}$ and $\hh{\x}$, so $\In{\x}$ is rotated into the cone determined by $\In{\x}$ and $\hh{\x}$.

Thus  ${\bf Out}(\x)= {\kappa_1} \In{\x} + {\kappa_2} \hh{\x}$ and ${\kappa_1}$, ${\kappa_2} >0$. In order for $[\x, {\bf Out}(\x)]$ to intersect $[\z, \hh{\x}]$ we must have that $\lambda \hh{\x} = \x +  \gamma{\bf Out}(\x)$ for $\gamma$, $\lambda>0$. That is,
\begin{align}
\lambda \hh{\x} & = \x +  \gamma{\bf Out}(\x)\\
                    & = \x + \gamma{\kappa_1} \In{\x} + \gamma{\kappa_2} \hh{\x},\\
                    & = -\In{\x} + \gamma{\kappa_1} \In{\x} + \gamma{\kappa_2} \hh{\x},\\
                    & = (\gamma{\kappa_1}-1) \In{\x} + \gamma{\kappa_2} \hh{\x}
\end{align}
Since $\In{\x}$ and $\hh{\x}$ are linearly independent, $\gamma{\kappa_1}-1=0$.
Thus $\gamma=1/{\kappa_1}>0$ and $\lambda = {\kappa_2}/{\kappa_1}>0$, establishing the physicality of $\x$. 

If ${\alpha_1}$, ${\alpha_2}<0$, then
$
\reflect{\In{\x}, \om{\x}} = \reflect{\In{\x}, -\om{\x}},
$
which gives the result.

In the other direction, say $\x \in \Phi_{C}$ is a physical point. Hence, there are $\lambda,\gamma>0$ with

\begin{align*}
\lambda\hh{\x}         & = \x + \gamma\Out{\x},\\
\lambda\hh{\x} + \In{\x}  & = \gamma\reflect{\In{\x}, \om{\x}},\\
\lambda\hh{\x} + \In{\x}  & = \gamma\left(-\In{\x} +2\frac{\In{\x}\cdot \om{\x}}{\om{\x}\cdot \om{\x}} \om{\x}\right),\\
\lambda\hh{\x} + (1+\gamma)\In{\x} & = 2\gamma\frac{\In{\x}\cdot \om{\x}}{\om{\x}\cdot \om{\x}} \om{\x},\\
\frac{\hh{\x}}{|\hh{\x}|}\lambda|\hh{\x}| + (1+\gamma)\In{\x}                                                  & =2\gamma\frac{\In{\x}\cdot \om{\x}}{\om{\x}\cdot \om{\x}} \om{\x}.\\ 
\end{align*}

Adding 
\[
0= \frac{|\hh{\x}|}{|\In{\x}|}\lambda\In{\x} - \frac{|\hh{\x}|}{|\In{\x}|}\lambda\In{\x}
\] 
to the left hand side and regrouping gives

\begin{align*}
|\hh{\x}| \lambda \mathscr{B}(\x) + \left[1+\gamma - \frac{|\hh{\x}| \lambda}{|\In{\x}|} \right] \In{\x}  = 2\gamma\frac{\In{\x}\cdot \om{\x}}{\om{\x}\cdot \om{\x}} \om{\x}.
\end{align*}

The positivity of the coefficient of $\In{\x}$ on the left hand side is  the form of the triangle inequality in lemma \ref{lem:triangle}. On the right hand side, the coefficient of $\om{\x}$ is positive because $\In{\x}\cdot \om{\x}$ is positive. Thus the coefficients of $\In{\x}$ and $\mathscr{B}(\x)$ are positive. 
\QED
\medskip

In practice, $\om{\x}$  usually has the form $\om{\x} = \omega_1(\x) \In{\x} + \omega_2(\x) \hh{\x}$, so we wish to express the above physicality condition in terms of $\omega_1$ and $\omega_2$. 
\begin{lemma}
Given $({\bf H}, \Omega)$,  
$\xz \in \br^3 - (\mathscr{E} \cup \mathscr{O}_2)$  is a physical point of $\Phi_{C}$ if and only if 
\[
\frac{\omega_1(\xz)}{\omega_2(\xz)} - \frac{\norm{\hh{\xz}} }{\norm{\In{\xz}}}>0.
\]
\end{lemma}

{\bf Proof.}
 Since $\In{\xz} \neq  \z$ and $\hh{\xz}\neq  \z$,
\begin{align}
\om{\xz} & = \omega_1 \In{\xz} + \omega_2 \hh{\xz}\\
           & = \alpha_1 \In{\xz} + \alpha_2 \mathscr{B}(\xz)\\
           & = \alpha_1 \In{\xz} + \alpha_2 \left( \frac{\In{\xz}}{\norm{\In{\xz}}} + \frac{\hh{\xz}}{\norm{\hh{\xz}}} \right)\\
           & = \left( \alpha_1 + \frac{\alpha_2}{\norm{\In{\xz}}} \right) \In{\xz} + 
                 \frac{\alpha_2}{\norm{\hh{\xz}}}\hh{\xz}                                           
\end{align}
Thus
\begin{align}
\omega_1 & = \alpha_1 + \frac{\alpha_2}{\norm{\In{\xz}}},\\
\omega_2 & = \frac{\alpha_2}{\norm{\hh{\xz}}}.
\end{align}
Therefore 
\begin{align}
\alpha_1 & = \omega_1 - \frac{ \omega_2\norm{\hh{\xz}}}{\norm{\In{\xz}}},\\ 
\alpha_2 & = \omega_2\norm{\hh{\xz}}, 
\end{align} 

giving our condition for the physicality of  $\x$ as 

\[
\alpha_1(\x)\alpha_2(\x) = \omega_2\norm{\hh{\xz}}  \left(\omega_1 - \frac{\omega_2\norm{\hh{\xz}} }{\norm{\In{\xz}}}\right) > 0.
\]
Thus $\x \in \br^3 - (\mathscr{E} \cup \mathscr{O}_2)$ is physical  iff

\begin{equation}
\frac{\omega_1(\x)}{\omega_2(\x)} - \frac{\norm{\hh{\xz}} }{\norm{\In{\xz}}}>0.
\label{eqn:physicaltest}
\end{equation}
\QED

Therefore, if $\omega_1 (\x_0)= 0$, then $\x_0$ is not a  physical point. 
On the other hand,  by definition $\mathscr{O}_2 \cup \mathscr{E}$ can't be physical. The above formula makes no sense on  $\mathscr{O}_2$, but it might for some elements of $\mathscr{E}$. Despite that, the points of $\mathscr{E}$ remain non-physical. One just needs to be careful when using equation \eqref{eqn:physicaltest}. Nevertheless, $\V(\x) = \In{\x}\times \hh{\x}$ is defined on all of $\br^3$, which has some interesting consequences as we will see in our examples below.

\begin{define} Given $({\bf H}, \Omega)$,
we define the \df{physicality function} 
\[
\rho: \br^3 -(\mathscr{E} \cup \mathscr{O}_2) \ra \br
\] 
as
\[
{\rho}(\x)  =   \frac{\omega_1(\x)}{\omega_2(\x)} - \frac{\norm{\hh{\x}} }{\norm{\In{\x}}}.
\]
\end{define}
Clearly $\rho$ is differentiable on $\br^3 - (\mathscr{E} \cup \mathscr{O}_2)$. Also, if $M \subseteq \Phi_C$ is an eigenmirror of ${\bf H}$, then $M \subseteq \rho^{-1}(0, \infty)$. 

\begin{lemma} \label{lem:nablaomega2perpV}
Given $({\bf H}, \Omega)$, for all $\x \in \mathscr{O}_2$,    
\[
\nabla \omega_2(\x) \cdot \V(\x)  =0.
\]
 
\end{lemma}
{\bf Proof.}  Since $\nabla \times [\omega_1(\x) \In{\x} + \omega_2(\x) \hh{\x}] = \z$, we have 
\[
\nabla \omega_1(\x)\times \In{\x} +
\omega_1(\x)\nabla \times \In{\x} +
\nabla \omega_2(\x) \times \hh{\x} +
\omega_2(\x) \nabla \times \hh{\x} =0.
\]
If $\x \in \mathscr{O}_2$, then $\omega_2(\x)=0$, so
\begin{equation}
\nabla \omega_1(\x)\times \In{\x} +
\nabla \omega_2(\x) \times \hh{\x}=0.
\label{eqn:o1Inom2H=0}
\end{equation}
Using \eqref{eqn:o1Inom2H=0}, we have
\begin{align}
|\nabla \omega_2(\x) \cdot [ \In{\x} \times \hh{\x}]| & = |\In{\x} \cdot  [ \nabla \omega_2(\x)  \times \hh{\x}]|,\\
                                                      & = |\In{\x} \cdot  [ -\nabla \omega_1(\x)  \times \In{\x}]|,\label{eqn:middle}\\
                                                      & = 0.
\end{align}
Thus $\nabla \omega_2(\x) \cdot [ \In{\x} \times \hh{\x}] = \nabla \omega_2(\x) \cdot \V(\x)  =0$ for $\x \in \mathscr{O}_2$. \QED
\medskip

\begin{lemma} 
\label{lem:maximalM}
Given $({\bf H}, \Omega)$, suppose that
\[
\widetilde{M} = {\Phi_C} \cap \rho^{-1}(0, \infty)
\]
is non-empty.
Then 
\begin{enumerate}[(A)]

\item $\widetilde{M}$ is a maximal eigenmirror.

\item If $\x_0 \in \br^3$ is a  boundary point of $\widetilde{M}$, then either (1) $\x_0 \in {\Phi_C}$ and  $\rho(\x_0)=0$  or (2) $\x_0 \in \mathscr{E} \cup \mathscr{O}_2$.
\end{enumerate}
 
\end{lemma}

{\bf Proof.}
(A)
 $\rho^{-1}(0, \infty)$ is open in $\br^3 -(\mathscr{E} \cup \mathscr{O}_2)$, and so, since $\Phi_C \subseteq \br^3 -(\mathscr{E} \cup \mathscr{O}_2)$ is a differentiable 2-manifold, it follows that $\widetilde{M} = \Phi_C \cap \rho^{-1}(0, \infty)$ is a differentiable 2-manifold. Additionally,  if $N\subseteq \Phi_C$ is an eigenmirror, then  $N \subseteq \rho^{-1}(0, \infty)$. Thus $N \subseteq {\Phi_C} \cap \rho^{-1}(0, \infty) = \widetilde{M}$, i.e., $\widetilde{M}$ is maximal.

(B) Say $\x_0$ is a boundary point of  $\widetilde{M}$. Then $\x_0$ lies in the closure of $\Phi_C$, and so  $\x_0 \in  {\Phi_C} \cup \mathscr{E} \cup \mathscr{O}_2$,  since ${\Phi_C} \cup \mathscr{E} \cup \mathscr{O}_2$ is closed.

If $\x_0 \in {\Phi_C}$ , then  either $\rho(\x_0)>0$,  $\rho(\x_0)<0$, or $\rho(\x_0)=0$. 

Suppose that $\rho(\x_0)>0$. There is a disk \( D \subseteq \Phi_C \), open in the topology of \( \Phi_C \), containing \( \x_0 \), on which \( \rho \) is positive.
 Thus $D$ contains points outside of $\widetilde{M}$ on $\Phi_C$, and $D \cup \widetilde{M}$ would be an eigenmirror. This  contradicts the maximality of $\widetilde{M}$, so $\rho(\x_0)>0$ is impossible.
 
If $\rho(\x_0)<0$, there on a disk $D$ about $\x_0$ on which  $\rho<0$, and so there would be points in $\widetilde{M}$ for which $\rho<0$, a contradiction. Thus $\rho(\x_0)=0$, because we know that $\rho$ is defined on $\Phi_C$. 

Of course, if $\x_0\not \in \Phi_C$, then $\x_0 \in \mathscr{E} \cup \mathscr{O}_2$, so the result follows. \QED

In example \ref{exam:twoboundaryloops}, all three types of points occur. 

\begin{lemma}
\label{lem:rhoofboundarypts}

Given $({\bf H}, \Omega)$, suppose that $\x_0 \in \Phi_C \subseteq \br^3 -(\mathscr{E} \cup \mathscr{O}_2)$ and $\rho(\x_0)=0$. Then $\Omega(\x_0)$ is a non-zero multiple of $\mathscr{B}$.
\end{lemma}

{\bf Proof.} For $\x_0 \in \br^3 -(\mathscr{E} \cup \mathscr{O}_2)$,
$\rho(\x_0)=0$ means that 
\[
\omega_1(\x_0)/ \omega_2(\x_0) = |\hh{\x_0}|/|\In{\x_0}|.
\] 
Since $\x_0 \in\br^3 -(\mathscr{E} \cup \mathscr{O}_2)$, $\hh{\x_0}\neq 0$, we have 
\[
\omega_2(\x_0) = \omega_1(\x_0)|\In{\x_0}|/|\hh{\x_0}|.
\]
Therefore,
\[
\Omega(\x_0) = \omega_1(\x_0) \In{\x_0} + \omega_1(\x_0)\frac{|\In{\x_0}|}{|\hh{\x_0}|} \hh{\x_0}.
\]
We cannot have that $\omega_1(\x_0)= 0$, because it implies that ${\bf H}(\x_0)=\z$, which is impossible since $\x_0 \in \br^3 -(\mathscr{E} \cup \mathscr{O}_2)$.
Thus, 
\[
\Omega(\x_0) = |\In{\x_0}|\omega_1(\x_0)\mathscr{B}(\x_0).
\]
\QED

\begin{figure}[htbp] 
\centering \includegraphics[height=1.6in]{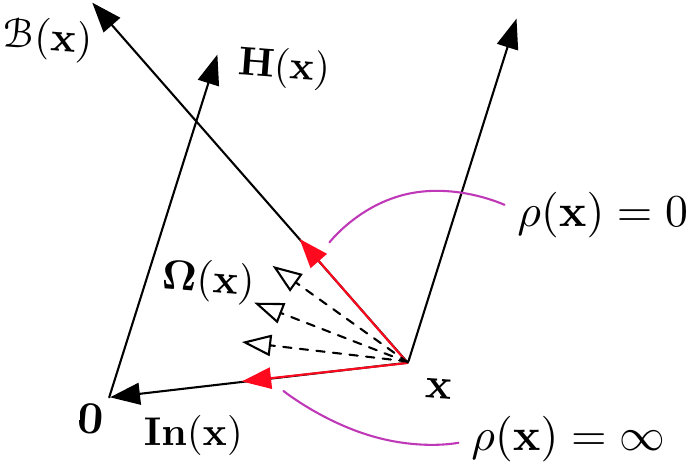} 
\caption{At a boundary point $\xz$ of a maximal eigenmirror $\widetilde{M}$, $\om{\xz}$ must be $\z$, or point in the direction of $\In{\xz}$ or $\mathscr{B}(\xz)$. We depict the latter two cases.}
\label{fig:two-ways-to-fail}
\end{figure}

\begin{lemma}
Given $({\bf H}, \Omega)$, suppose that $\widetilde{M}$ is a non-empty maximal eigenmirror of $\Phi_C$ as in lemma \ref{lem:maximalM}, and that $\x_0$ is a boundary point of $\widetilde{M}$. Then $\Omega(\x_0)$ is a non-zero multiple of $\mathscr{B}(\x_0)$ or a multiple of $\In{\x_0}$. (See Fig. \ref{fig:two-ways-to-fail}.)
\end{lemma}

{\bf Proof.} We know that $\x_0 \in {\Phi_C} \cup \mathscr{E} \cup \mathscr{O}_2$.

If $\x_0 \in {\Phi_C}$, then by lemma \ref{lem:maximalM},
$\rho(\x_0)=0$. Then by  lemma \ref{lem:rhoofboundarypts}, $\Omega(\x_0)$ is a non-zero multiple of  $\mathscr{B}(\x_0)$.

If $\x_0 \in \mathscr{E}$,  then $\hh{\x_0}=s\x_0$ for some $s\in \br$. Thus
\[
\om{\x_0} = \omega_1(\x_0)\In{\x_0} + \omega_2(\x_0)s\x_0 =  
(\omega_1(\x_0) - \omega_2(\x_0)s)\In{\x_0}.
\] 

If $\x_0 \in \mathscr{O}_2$, then $\omega_2(\x_0)=0$, so  $\om{\x_0} = \omega_1(\x_0)\In{\x_0}$. 

\QED 

The above tells us that there are threes ways to ``leave'' an eigenmirror - one must cross a boundary point where $\Omega$ points towards the origin, or $\Omega$ points in the direction of $\mathscr{B}$, or $\Omega$ vanishes. We will see below that in the first case, integral curves of $\V$ will be repelled.

\begin{lemma} 
\label{lem:nocontact}
Given $({\bf H}, \Omega)$, suppose $\nabla \omega_2$ does not vanish on $\mathscr{O}_2$. 
Then an integral curve  of $\V$ either lies in $\mathscr{O}_2$ or never intersects it. 
\end{lemma}

{\bf Proof.} In brief, $\mathscr{O}_2$ is a manifold tangent to a non-vanishing vector field, so an integral curve starting off of $\mathscr{O}_2$ cannot intersect it, or the uniqueness of integral curves is violated. \QED



\begin{theorem} 
\label{thm:trapped}
Given $({\bf H}, \Omega)$, suppose $\widetilde{M}$ is a non-empty maximal  eigenmirror of $\Phi_C$, that $\partial \widetilde{M} \subseteq \mathscr{O}_2$ and that $\nabla \omega_2$ does not vanish. Then no integral curve of $\V$ starting in $\widetilde{M}$ contains a point of $\partial \widetilde{M}$.
\end{theorem}

{\bf Proof.} Let $\chi:(a,b)\ra \Phi_C$ be  an integral curve of $\V$, with $a<t_0 <t_1 <b$, $\chi(t_0) \in  \widetilde{M}$ and $\chi(t_1) \in \partial \widetilde{M}$. Then $\chi(t_1)  \in \mathscr{O}_2$. But by lemma \ref{lem:nocontact} the trace of $\chi$ cannot intersect $\mathscr{O}_2$. \QED

This is not an uncommon situation, as we will see in our examples.
\section{More Examples}

In these examples, we are going to work backwards to find examples. Rather than starting with given ${\bf H}$, we are going to treat ${\bf H}$, $\omega_1$, $\omega_2$ as symbolic entities (unknowns). These computations were done with a symbolic computing system that could solve our polynomial systems in a short time, but the resulting output was often challenging to manage by a human. The main equation was $\curlnp \Omega = \z$. Generally when solving such a polynomial system we obtained large number of solutions, and what is presented here barely scratches the surface of the results. For the first example below, 81 solutions were found. Excluding complex solutions reduced this to 60, and then excluding singular solutions gave 40, and lastly excluding complex solutions gave 21. Each of these had free parameters that needed assignments. While these parameters were free, it was easy to create divisions by zero. This was  avoided by assigning values that were ratios like $a_{3} = {\frac{434}{729}}$. The only drawback to this was the awkward appearance on paper.

It was also fairly easy to create an $\Omega$ that contained a rich collection of eigenmirrors simply by using the fact that
\[
{\rho}(\x)  =   \frac{\omega_1(\x)}{\omega_2(\x)} - \frac{\norm{\hh{\x}} }{\norm{\In{\x}}}.
\]

and then assign values to $\omega_1$ that were an order of magnitude bigger than $\omega_2$.

Since we can't reasonably include all of the data associated with the below examples,  any missing information here or elsewhere is available upon request.

\begin{example}
\label{exam:twoboundaryloops}
In this example, $M$ lies on a  compact connected component of a real algebraic variety of degree 3.
Let's assume that ${\bf H}$ is linear with matrix
\[
H = \left[\begin{array}{ccc}
a_{1} & a_{2} & a_{3} 
\\
 b_{1} & b_{2} & b_{3} 
\\
 c_{1} & c_{2} & c_{3} 
\end{array}\right]
\]
and 
$
\omega_1 = \mu_{4}x + \mu_{3}y + \mu_{2}z+\mu_{1}
$
and
$
\omega_2 = \nu_{10}x^{2} +\nu_{9}x y +  \nu_{8}x z + \nu_{6}y^{2} + \nu_{5}y z  + \nu_{3}z^{2} + \nu_{7}x +\nu_{4}y  +\nu_{2}z  +\nu_{1}
$. 
That is, ${\bf H}$, $\omega_1$, and $\omega_2$ contain symbolic values. 
We then considered the equation $\curlnp \Omega = \z$ and applied a symbolic solver. This resulted in 81 solutions which are parametrized by some of the symbolic values from above. One solution is 
\begin{dmath}
\left\{a_{1} = 
\left(2 a_{3}^{2} b_{3} \nu_{4}^{2}-4 a_{3}^{2} c_{2} \nu_{4}^{2}+7 a_{3} b_{1} b_{3} \nu_{2} \nu_{4}-2 a_{3} b_{1} c_{2} \nu_{2} \nu_{4}-4 b_{1}^{2} b_{3} \nu_{2}^{2}+2 b_{1}^{2} c_{2} \nu_{2}^{2}-\\
18 b_{3}^{3} \nu_{2}^{2}+33 b_{3}^{2} c_{2} \nu_{2}^{2}+9 b_{3}^{2} c_{3} \nu_{2} \nu_{4}-20 b_{3} c_{2}^{2} \nu_{2}^{2}-12 b_{3} c_{2} c_{3} \nu_{2} \nu_{4}+4 c_{2}^{3} \nu_{2}^{2}+4 c_{2}^{2} c_{3} \nu_{2} \nu_{4}\right)/\\
\left({\nu_{2} \left(3 b_{3}-2 c_{2}\right)^{2} \nu_{4}}\right)
,\\ a_{2} = 
\frac{-a_{3} b_{3} \nu_{4}+2 a_{3} c_{2} \nu_{4}+2 b_{1} b_{3} \nu_{2}-2 b_{1} c_{2} \nu_{2}}{\nu_{2} \left(3 b_{3}-2 c_{2}\right)}
,\\ 
b_{2} = -\frac{2 b_{3} \nu_{2}^{2}+b_{3} \nu_{4}^{2}-c_{2} \nu_{2}^{2}-2 c_{2} \nu_{4}^{2}-c_{3} \nu_{2} \nu_{4}}{\nu_{2} \nu_{4}}
,\\ c_{1} = 
\frac{2 a_{3} b_{3} \nu_{4}-2 a_{3} c_{2} \nu_{4}+2 b_{1} b_{3} \nu_{2}-b_{1} c_{2} \nu_{2}}{\left(3 b_{3}-2 c_{2}\right) \nu_{4}}
,\\  \mu_{2} = 
-\frac{\left(2 b_{3} \nu_{2}-c_{2} \nu_{2}-c_{3} \nu_{4}\right) \nu_{2}}{\nu_{4}}
, \mu_{3} = -2 b_{3} \nu_{2}+c_{2} \nu_{2}+c_{3} \nu_{4},\\ 
\mu_{4} = 
\frac{-4 a_{3} b_{3} \nu_{2} \nu_{4}+2 a_{3} c_{2} \nu_{2} \nu_{4}+2 a_{3} c_{3} \nu_{4}^{2}+2 b_{1} b_{3} \nu_{2}^{2}-b_{1} c_{2} \nu_{2}^{2}-b_{1} c_{3} \nu_{2} \nu_{4}}{\nu_{4} \left(3 b_{3}-2 c_{2}\right)}
,\\ \nu_{1} = 0, \nu_{3} = 0,  
\nu_{5} = 0, \nu_{6} = 0, \nu_{7} = 
\frac{2 a_{3} \nu_{4}-b_{1} \nu_{2}}{3 b_{3}-2 c_{2}}, \nu_{8} = 0, 
\nu_{9} = 0, \nu_{10} = 0\right\}
\end{dmath}
Any of the above parameters that appear on a right hand side are free, so in  a quasirandom manner (and trying to avoid later divisions by zero) we take
\[
a_{3} = {
\frac{434}{729}}, b_{1} = {\frac{437}{243}},  b_{3} = -{\frac{362}{243}},  c_{2} = -{
\frac{362}{243}}, c_{3} = -{\frac{395}{243}}, \mu_{1} = {\frac{584}{
243}},   \cdots 
\]

We intentionally took these values to be rational, since for many of our question we desire exact answers. ${\bf H}$ is 
\[
\left[
-\frac{38638834374205 x}{3775648458897}-\frac{1546342 y}{1951533}+\frac{434 z}{729}
, \frac{437 x}{243}-\frac{1766201965 y}{2317770693}-\frac{362 z}{243}
, -\frac{1169849 x}{865809}-\frac{362 y}{243}-\frac{395 z}{243}\right]
\]
Our homogeneous quadratic vector field is
\begin{dmath*}
\bigg[
-y \left(-\frac{1169849 x}{865809}-\frac{362 y}{243}-\frac{395 z}{243}\right)+z \left(\frac{437 x}{243}-\frac{1766201965 y}{2317770693}-\frac{362 z}{243}\right)
, 
\end{dmath*}
\begin{dmath*}
x \left(-\frac{1169849 x}{865809}-\frac{362 y}{243}-\frac{395 z}{243}\right)-z \left(-\frac{38638834374205 x}{3775648458897}-\frac{1546342 y}{1951533}+\frac{434 z}{729}\right)
, 
\end{dmath*}
\begin{dmath*}
-x \left(\frac{437 x}{243}-\frac{1766201965 y}{2317770693}-\frac{362 z}{243}\right)+y \left(-\frac{38638834374205 x}{3775648458897}-\frac{1546342 y}{1951533}+\frac{434 z}{729}\right)
\bigg]
\end{dmath*}
Also,
\[
\omega_1 = \frac{193702855309 x}{152323508988}-\frac{29339 y}{39366}+\frac{78540503 z}{140261058}+\frac{584}{243}
\]
and 
\[
\omega_2 =-\frac{6602231 x}{14250492}+\frac{3563 y}{13122}-\frac{2677 z}{13122}.
\]
Thus $\nabla \omega_2$ never vanished.
Since   $\curlnp{\Omega} = \z$, it is not hard to compute
\begin{dmath}
\phi = \frac{126530246968627014551}{161414544474972081972}+\frac{93340670007779573599 x^{3}}{80707272237486040986}+\frac{\left(-14826154874832-10279878336161 y +7723613333119 z \right) x^{2}}{12338204228028}-\frac{217 \left(3563 y -2677 z \right)^{2} x}{12804008013}+\frac{2297789389 y^{3}}{12804008013}+\frac{\left(-1915812-644903 z \right) y^{2}}{1594323}+\frac{484537 y \,z^{2}}{1594323}-\frac{1297105549 z^{3}}{17041718547}-\frac{292 z^{2}}{243}
.
\end{dmath} 

Observe that $\phi$ is cubic, although not homogeneous ($\omega_1$ is not). We then choose $C=\phi(1/2,1/2,1/2)$ - see Fig. \ref{fig:phi1}A. In Fig. \ref{fig:phi1}B we see  $\Phi_C$ along with  the surfaces $\rho(x,y,z)=0$ and $\mathscr{O}_2$. In Fig. \ref{fig:phi1}C we have hatched the eigenmirror, $M$.  Let $\ell_1$ denote the upper loop of $\partial M$, where $\rho(x,y,z)=0$, and $\ell_2 = \mathscr{O}_2 \cap M$, the lower loop.
\newcommand\sizec{2.0}
\begin{figure}[htbp] 
\begin{center}
A. \includegraphics[height=\sizec in]{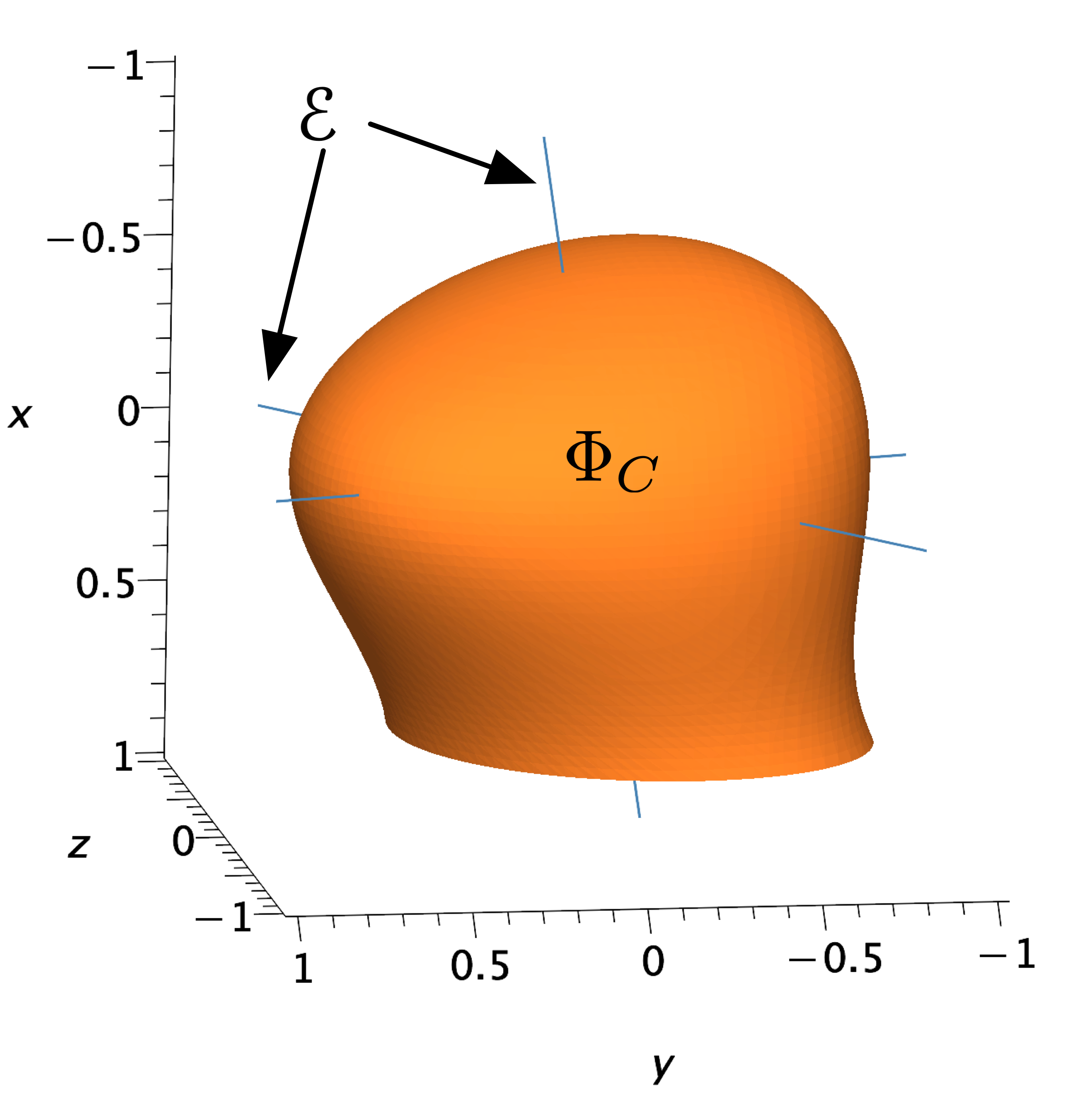} 
B. \includegraphics[height=\sizec in]{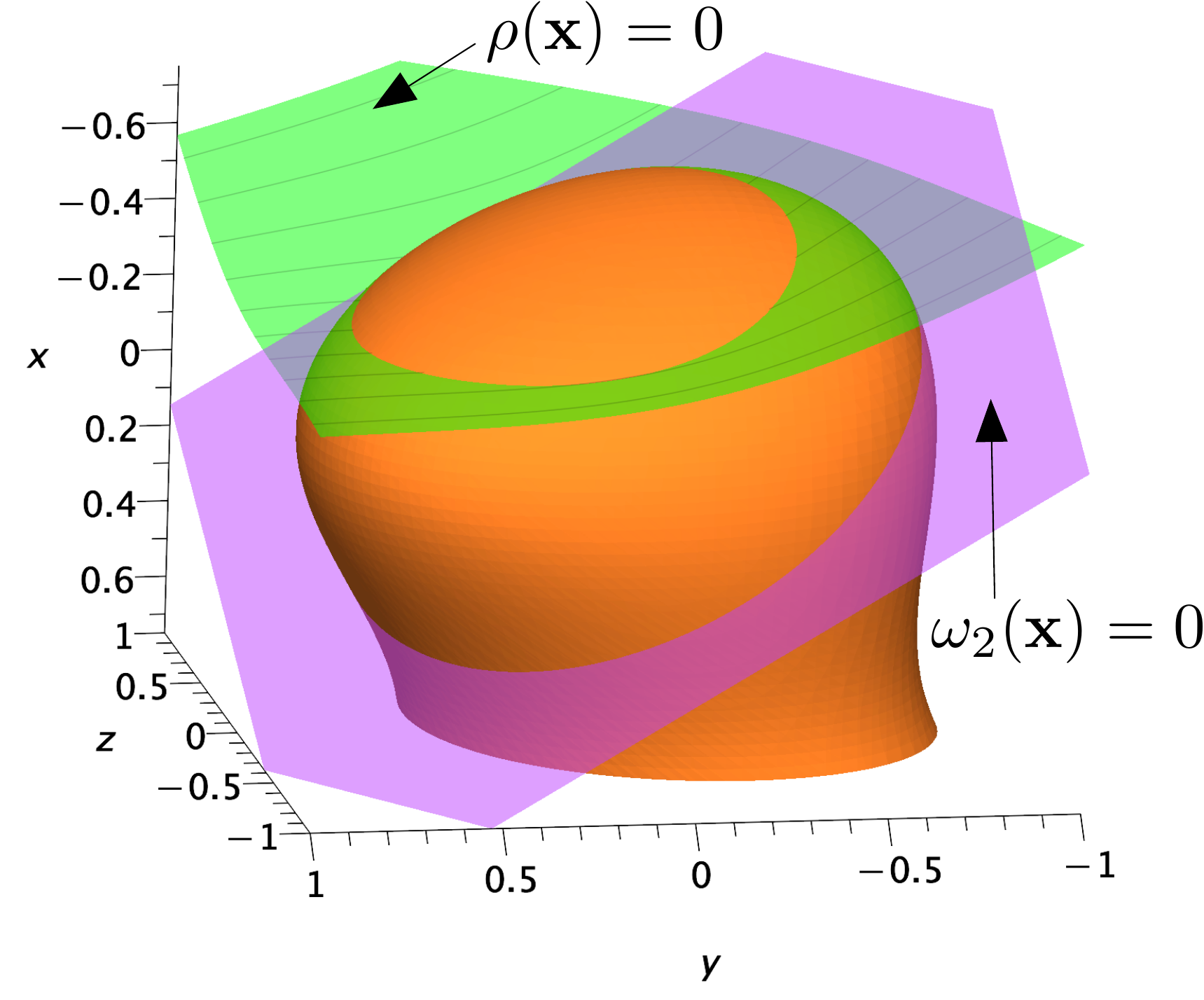}\\
C. \includegraphics[height=\sizec in]{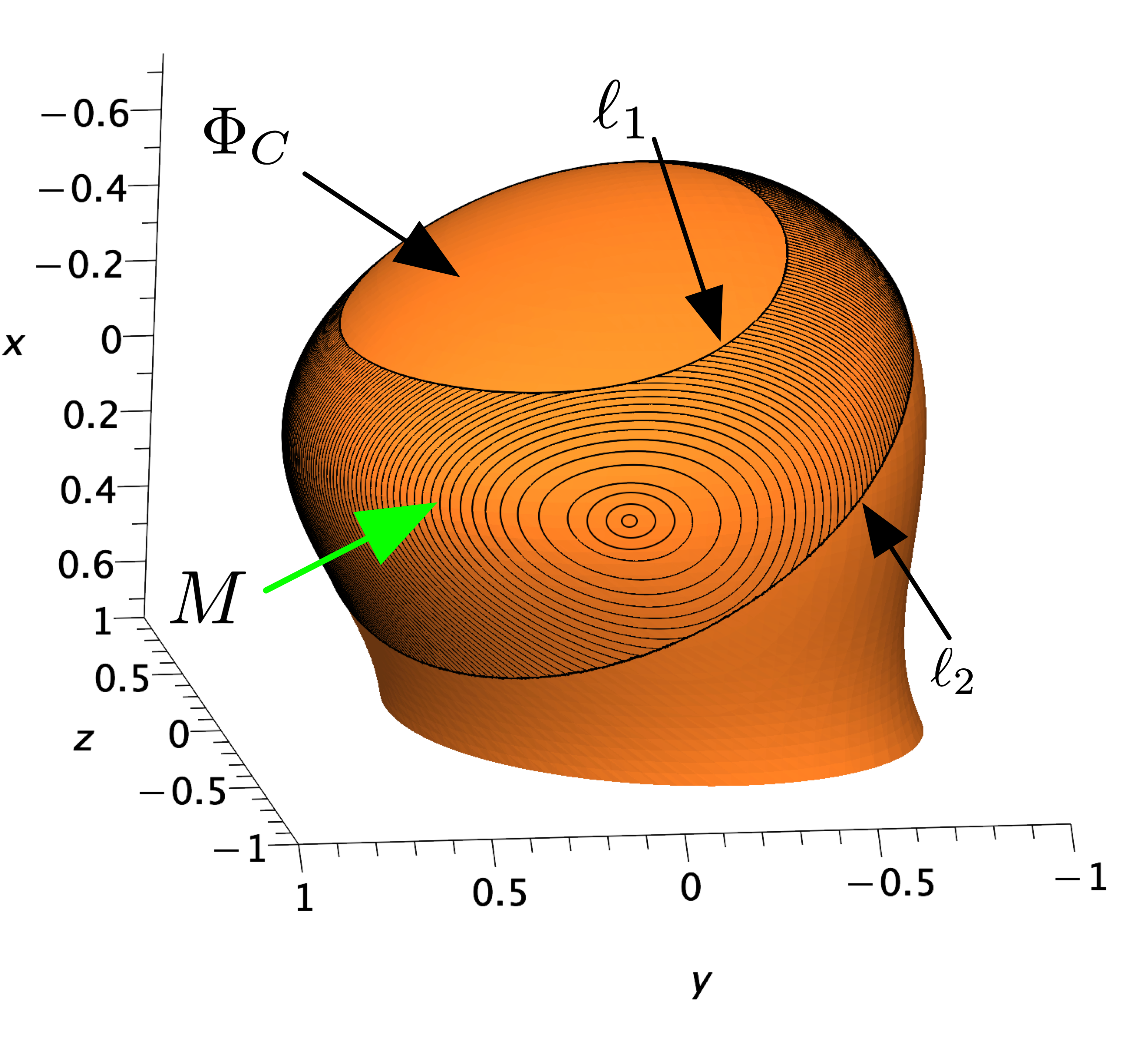}
D. \includegraphics[height=\sizec in]{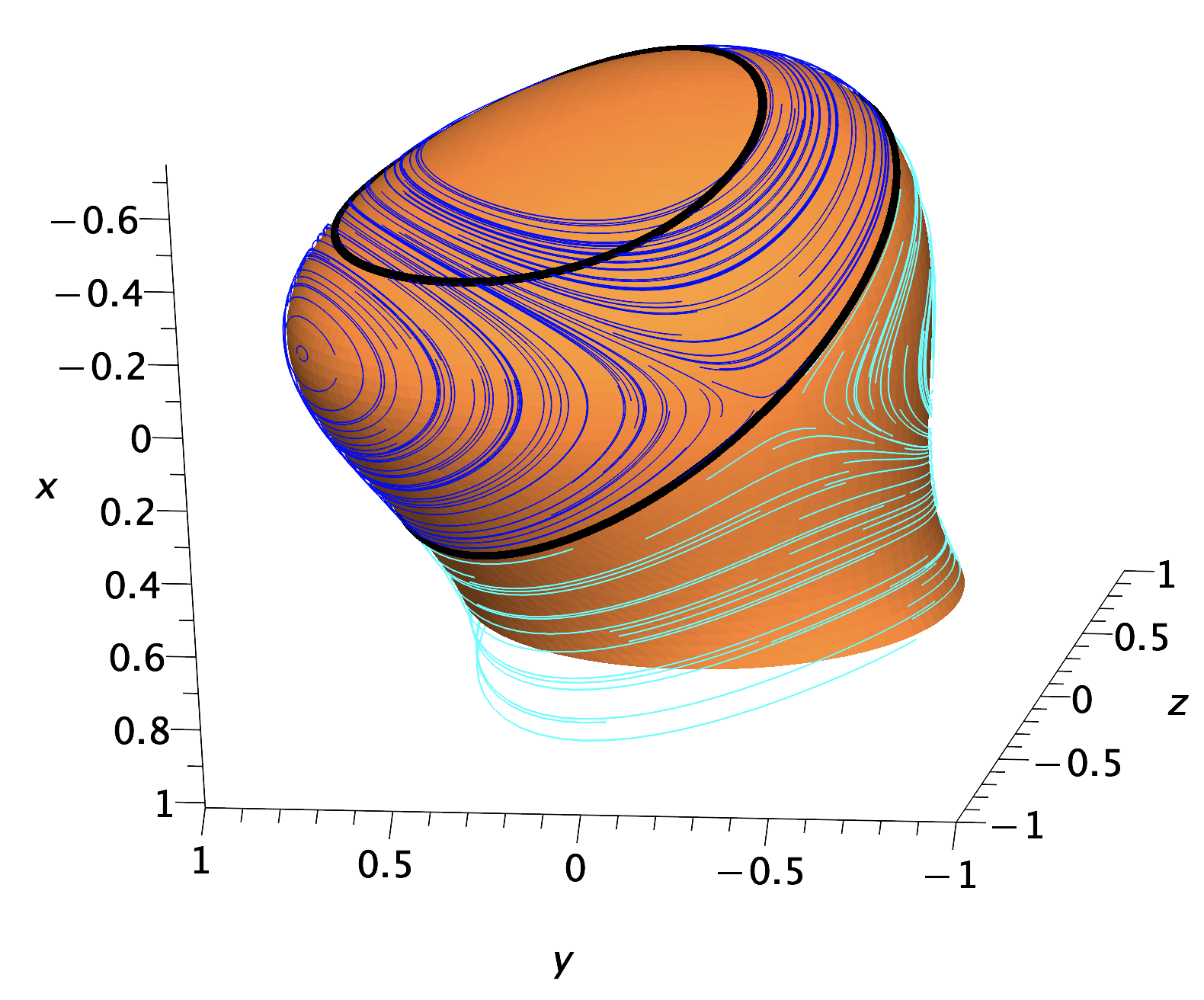}
\end{center}
\caption{A. A portion of $\Phi_C$ and $\mathscr{E}$. B. $\Phi_C$, the surface $\rho(\x)=0$ in green, and $\mathscr{O}_2$ in violet. C. $\Phi_C$, where $M$ is the  blacked hatched section bounded by the loops $\ell_1$ and $\ell_2$.  D. Some integral curves of $\dot{\x} = \In{\x}\times \hh{\x}$. They may cross $\ell_1$, since there  $\rho(\x)=0$, but not $\ell_2$.}
\label{fig:phi1}
\end{figure}
In Fig. \ref{fig:phi1}D we see some integral curves of $\V$ on  $\Phi_C$, which don't cross $\ell_2$, as was argued in lemma \ref{lem:nocontact}. No such argument can be made for $\ell_1$. Additionally,  for $\x \in M$ near $\ell_1$, $\Omega{\x}$ is pointing nearly in the direction of $\mathscr{B}(\x)$, and so  $\Out{\x}$ is nearly parallel to $\hh{\x}$. Consequently the intersection of $\Out{\x}$ and $\hh{\x}$ is far away, and hence $S$ is unbounded. Indeed, in this example, which comprises a finite number of points, the diameter is large.
\begin{figure}[htbp] 
\begin{center}
\includegraphics[height=1.85in]{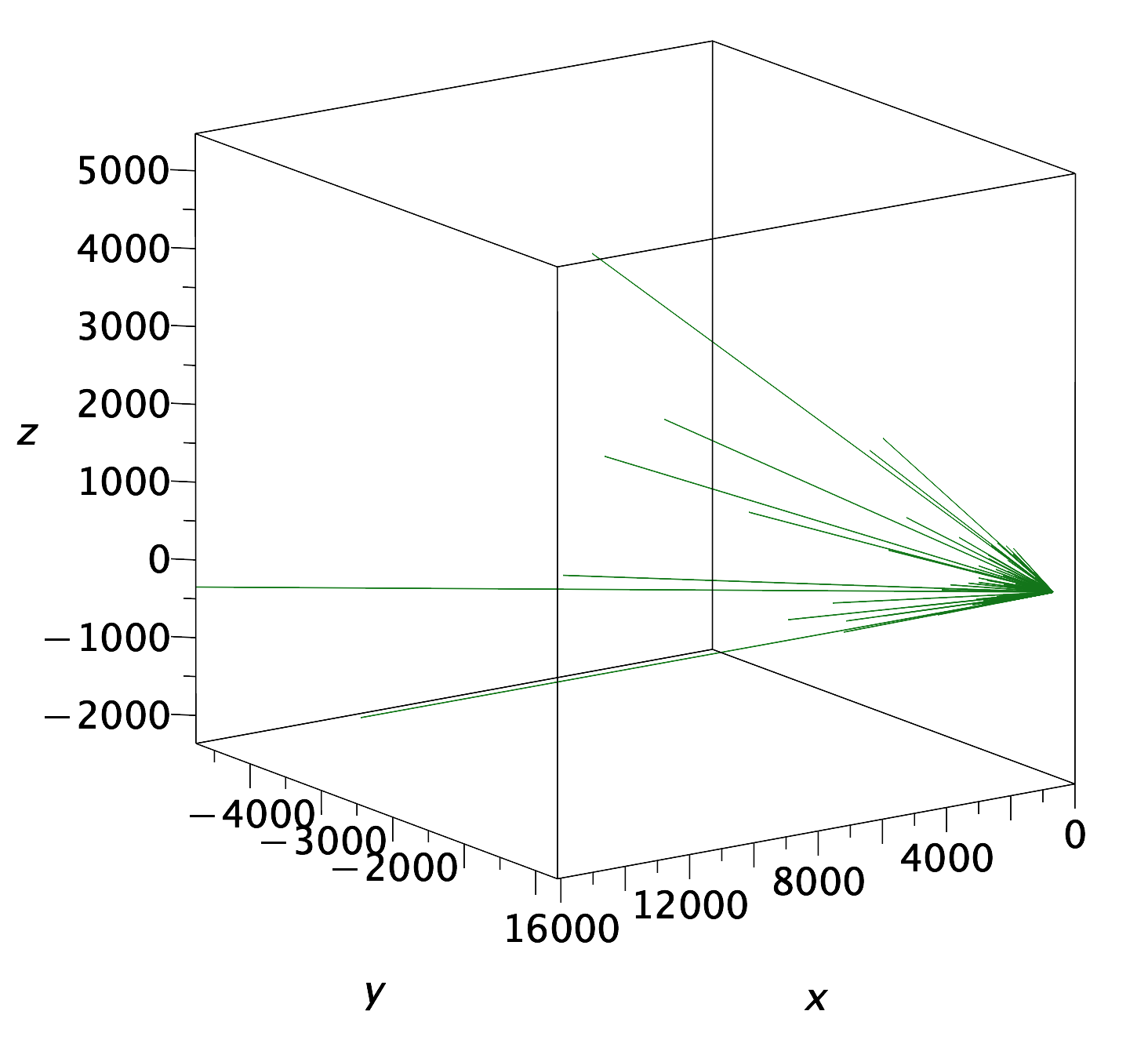} 
\caption{The eigensurface of example \ref{exam:twoboundaryloops}. }
\end{center}
\label{fig:es-ex1}
\end{figure}

\end{example}

\begin{example}
\newcommand\sizea{2.1} 
\begin{figure}[htbp] \centering
\includegraphics[width=\sizea in]{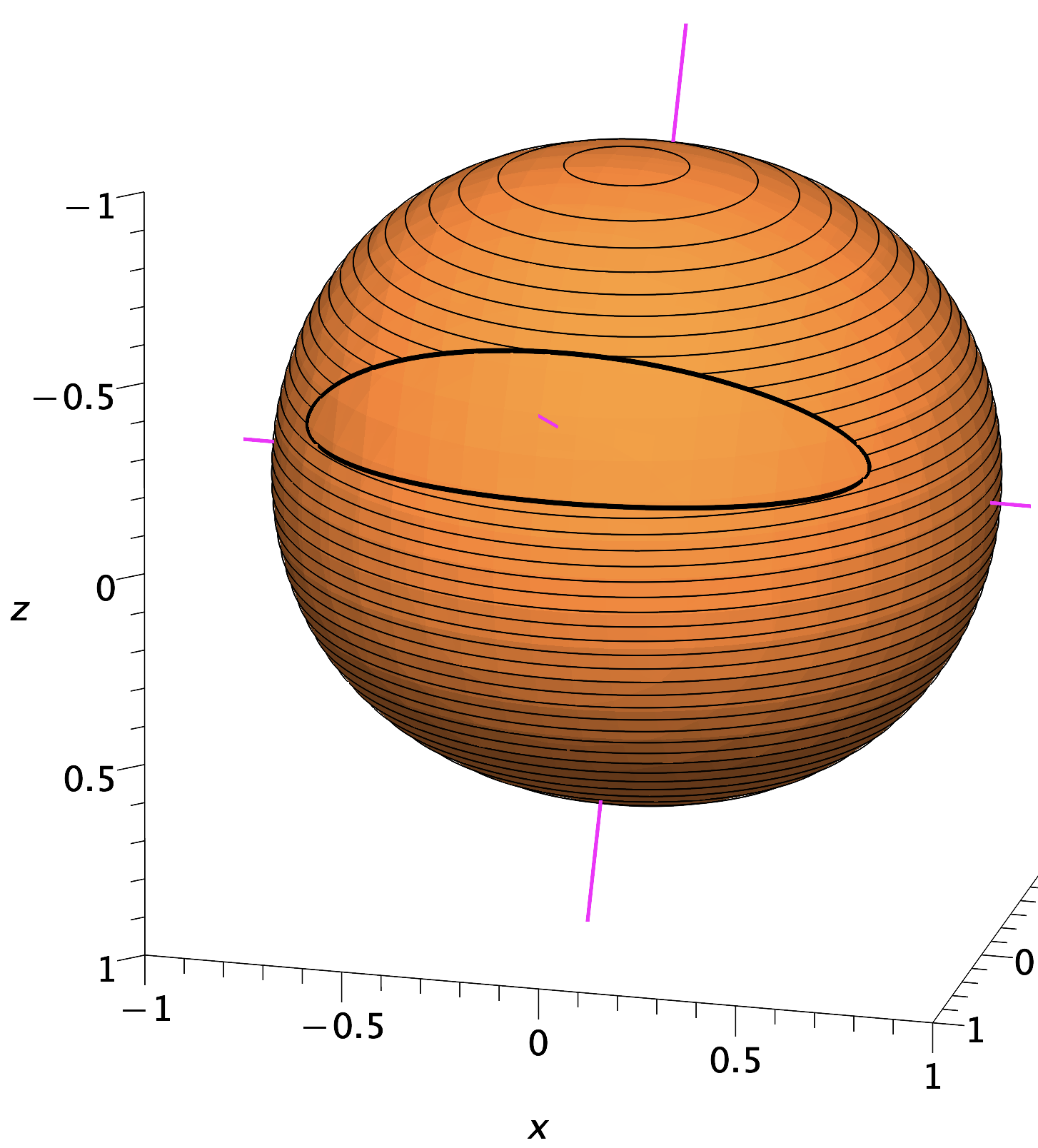} 
\includegraphics[width=\sizea in]{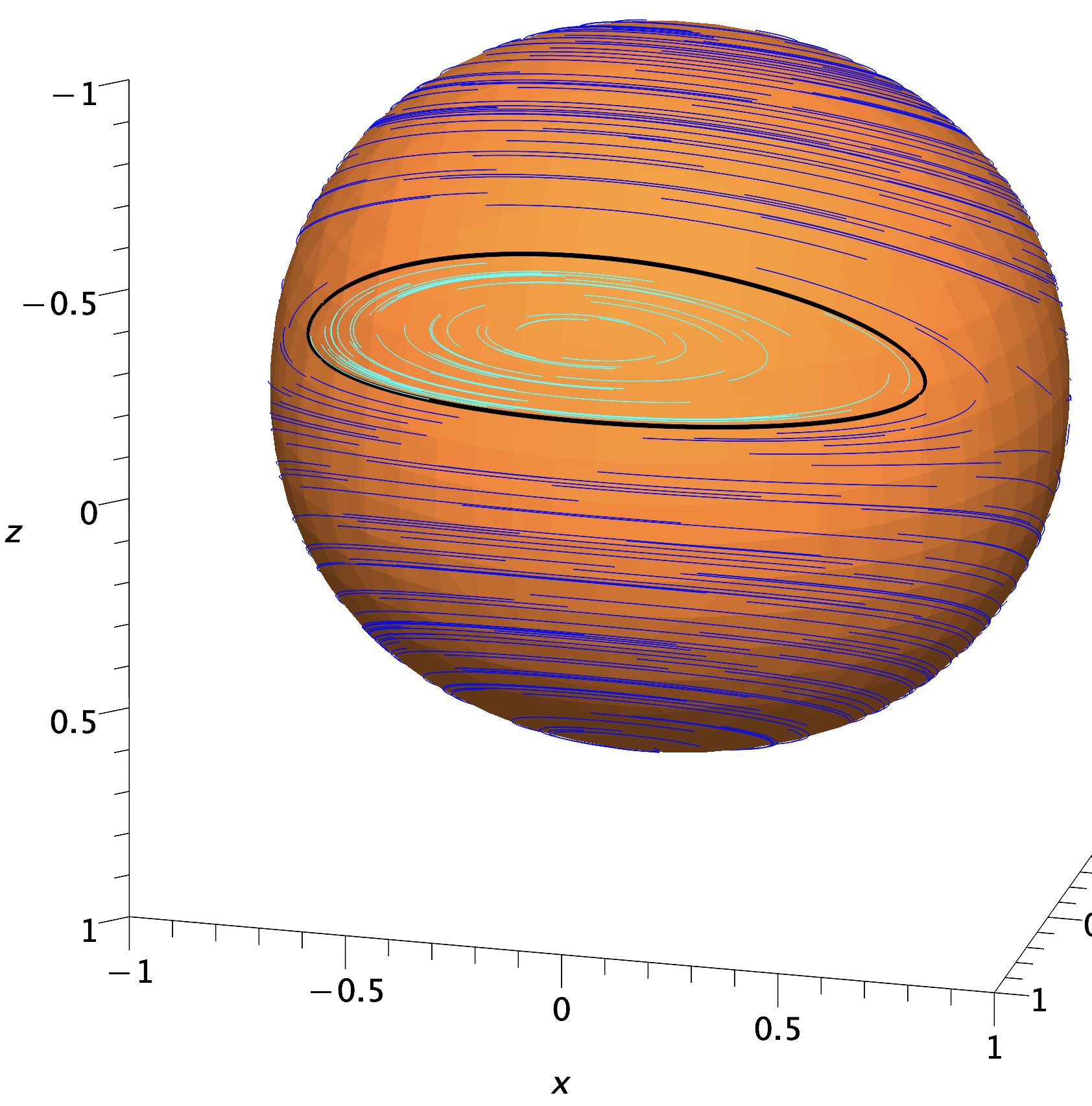}
\caption{On the left we see a connected component of $\Phi_C$. The black hatched region represents the eigenmirror $M$, and the magenta lines the eigenspaces of ${\bf H}$. The black curve that bounds the oviform region is where $\omega_2(\x)=0$. (There is another such oviform on the back side.) On the right we see $\Phi_C$, superimposed with the orbits of $\dot{\x}=\V(\x)$. Orbits may not cross the black curve, and are colored blue if the start in $M$ and cyan otherwise. Thus this connected component of $M$ is invariant under the flow of $\V$. }
\label{fig:twoeyes}
\end{figure}

\newcommand\sizeb{2.1}
\begin{figure}[htbp] \centering
\includegraphics[width=\sizeb in]{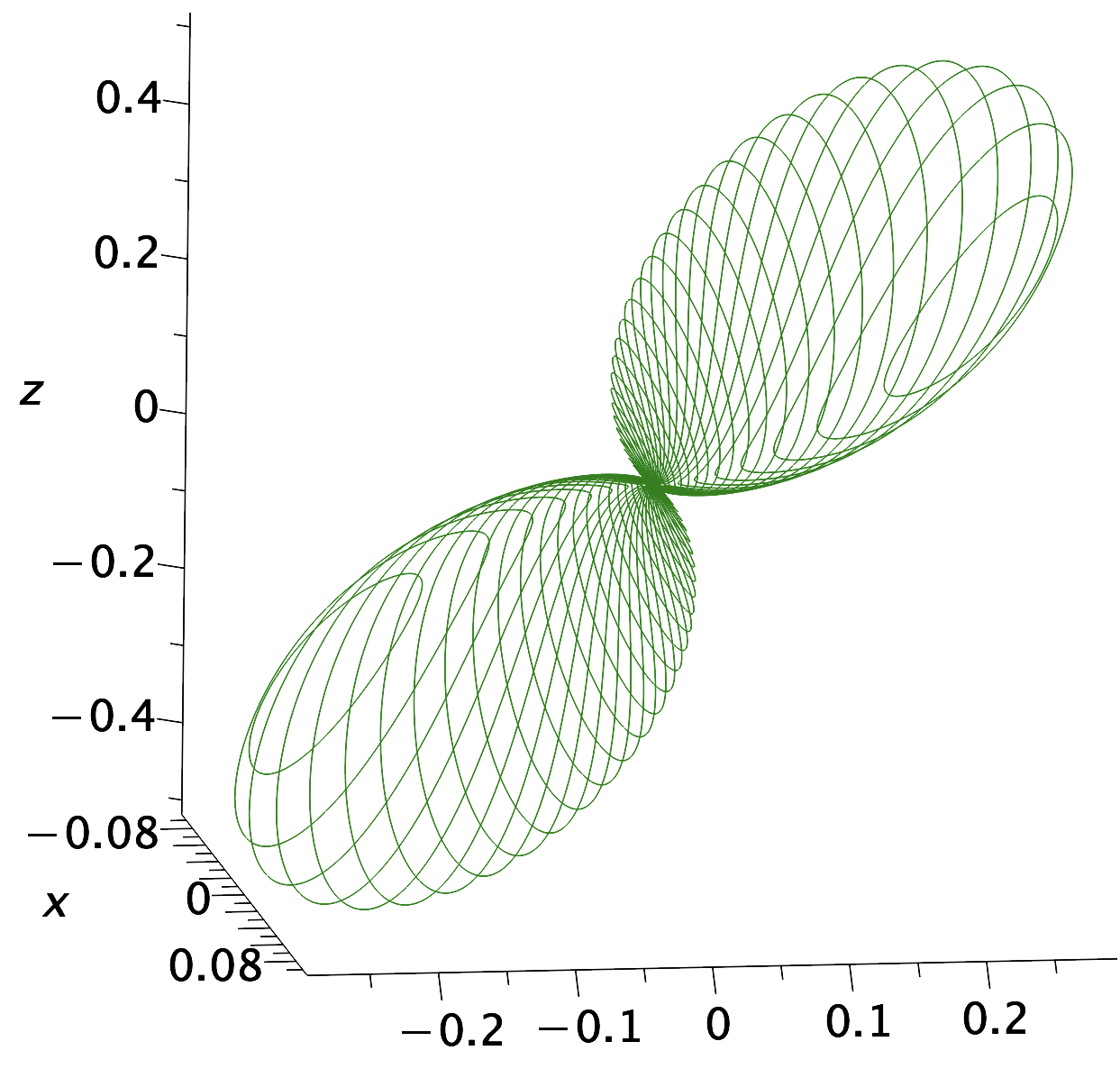}
\includegraphics[width=\sizeb in]{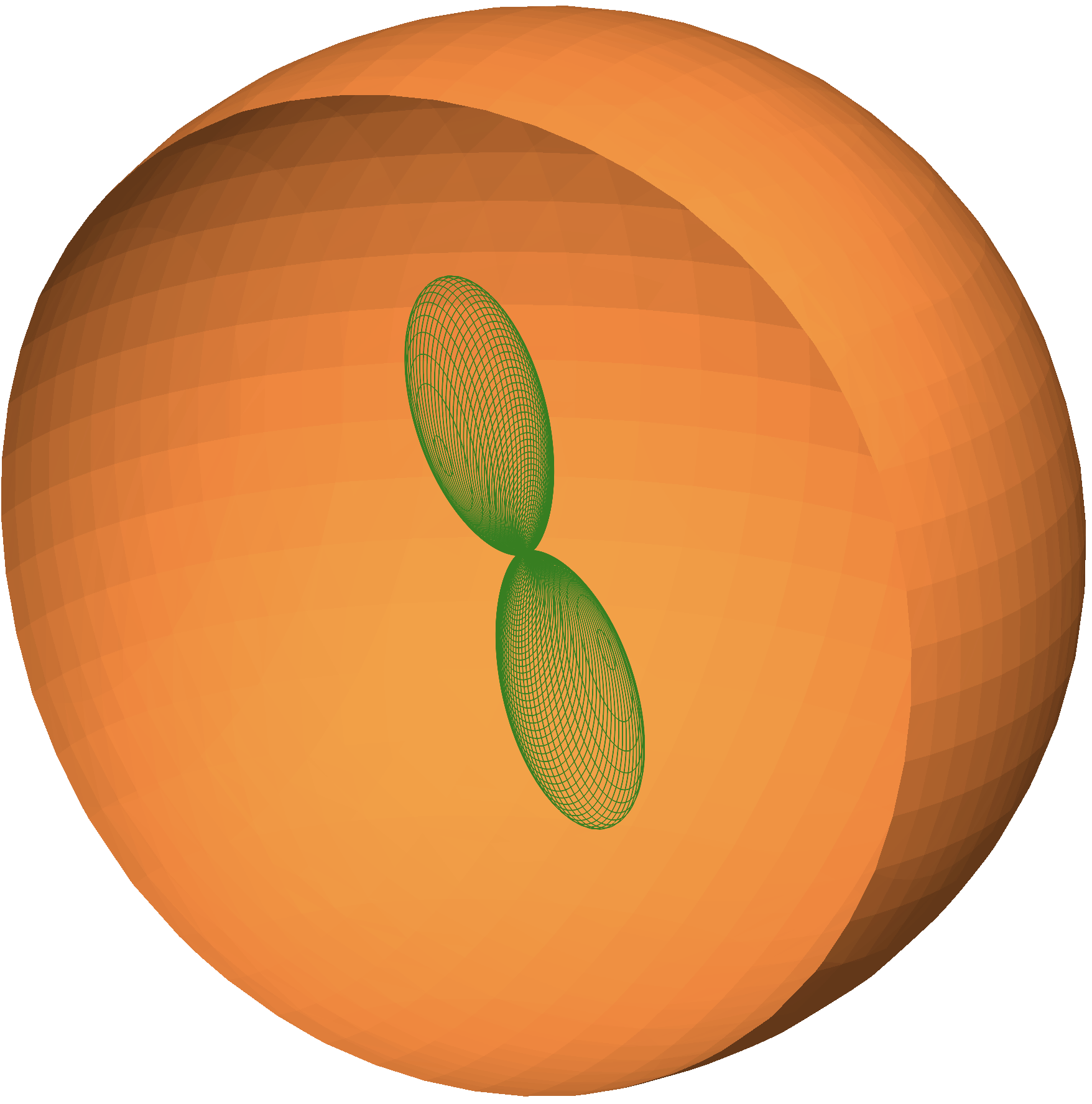}
\caption{On the left we have the eigensurface of $M$. On the right we see a cutaway of $M$ and $S$ plotted together. }
\label{fig:SandM}
\end{figure}

In this example, $M$ lies on a  compact connected component of a real algebraic variety of degree 4. In this example, $\partial M \subseteq \mathscr{O}_2$,  so by theorem \ref{thm:trapped},  $M$ is invariant under the flow of $\V$.  We made same symbolic choices as in example \ref{exam:twoboundaryloops}, obtaining 
\[
{\bf H} = \left[\frac{265 x}{243}, \frac{434 y}{729}-\frac{362 z}{243}, 
-\frac{362 y}{243}-\frac{395 z}{243}.\right]
\]
We found (again using some quasirandom assignments) that one solution is

\begin{align*}
\phi & = -\frac{\left(846702825 x^{2}+462225190 y^{2}-2313256020 y z -1262066475 z^{2}+15898047396\right)^{2}}{23456171485373565600}\\
     &  -\frac{105412 y^{2}}{193185}-\frac{211408 y z}{64395}-\frac{12848 z^{2}}{4293}+\frac{8595450631421967263}{733005358917923925}.
\end{align*}
In this case  $\nabla \omega_2$ vanishes only at the origin.
Unlike our first example, here $S$ is bounded because, since $\Omega$ is bounded away from $\mathscr{B}$ on $M$.
\end{example}

\begin{example} 
Taking 

$
{\bf H}=
$

$
[x^{2} a_{4}+x y a_{1}+x z a_{2}+y^{2} a_{5}+y z a_{3}+z^{2} a_{6},
$

$
x^{2} b_{4}+x y b_{1}+x z b_{2}+y^{2} b_{5}+y z b_{3}+z^{2} b_{6}, 
$

$
x^{2} c_{4}+x y c_{1}+x z c_{2}+y^{2} c_{5}+y z c_{3}+z^{2} c_{6}]
$

and $\omega_1$, $\omega_2$ of degree 1 and 2 as above, we obtain 32 solutions. One of which leads to
 
the degree five potential 
\begin{dmath}
\scalemath{.9}{\phi = \frac{\left(-362416263729113905 y^{2}+595647951606298788 y z +147316161615536505 z^{2}\right) x^{3}}{1241971345424369484}+\cdots} 
\end{dmath}
 A portion of the $\Phi_C$ passing through $(1/2,1/2,1/2)$  appears in Fig. \ref{fig:SandMquadratic}. Again, $\partial M \subseteq \mathscr{O}_2$, so integral curves starting in $M$ are trapped in $M$.
\end{example}

\newcommand\sizex{2.25}
\begin{figure}[htbp] \centering
\includegraphics[width=\sizex in]{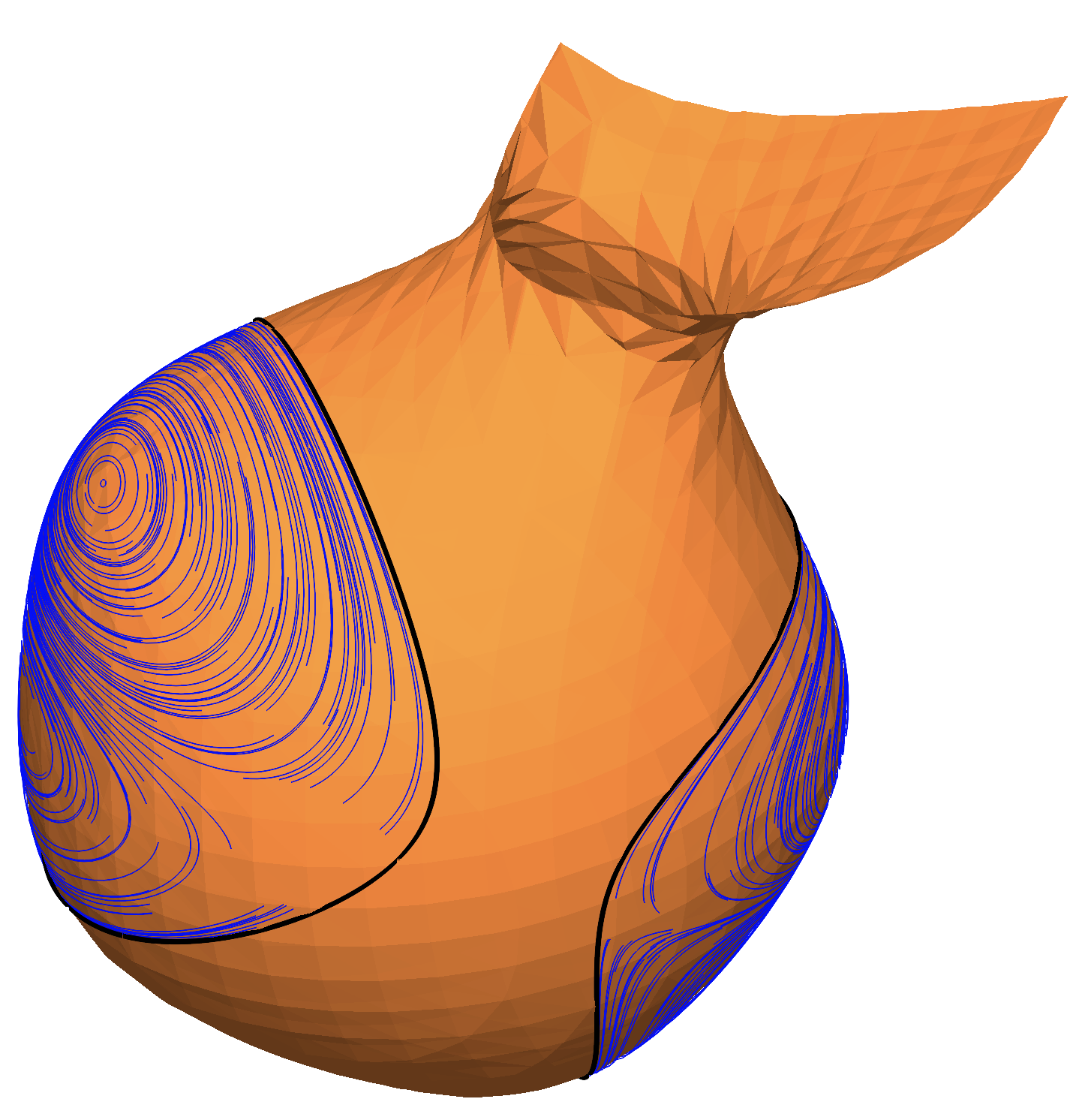}
\includegraphics[width=\sizex in]{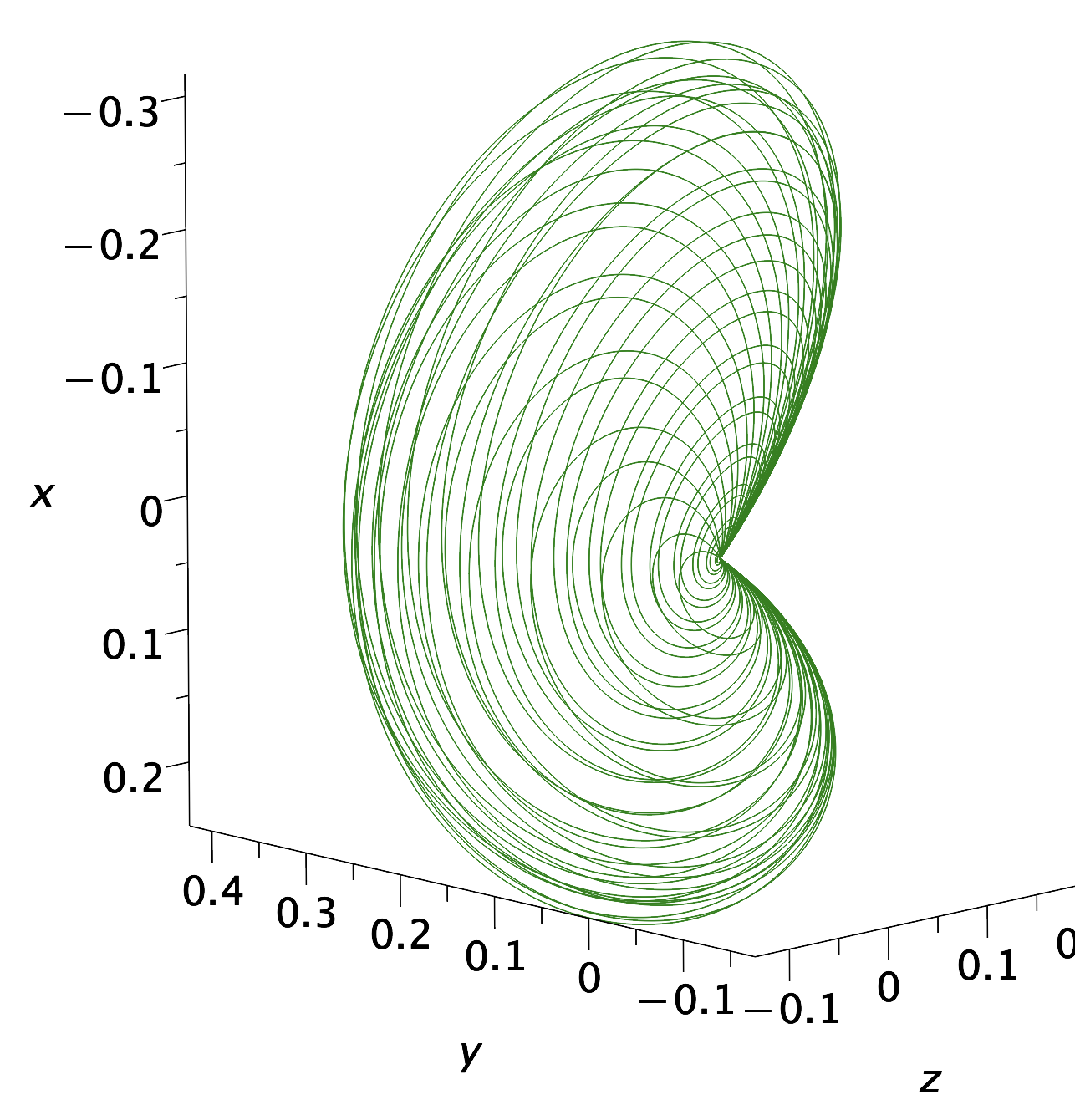}
\caption{On the left we see some blue integral curves in a degree five eigenmirror. As earlier, the black curves forming the repelling boundary are part of $\Phi_C \cap \mathscr{O}_2$. On the right we see the corresponding eigensurface in the case of a quadratic ${\bf H}$. $S$ is close to being flat.}
\label{fig:SandMquadratic}
\end{figure}


\bibliographystyle{sn-mathphys-num}

\end{document}